\documentclass[aip,jcp,reprint,floatfix]{revtex4-1}
\usepackage[utf8]{inputenc}
\usepackage{amsmath}
\usepackage{amsfonts}
\usepackage{amssymb}
\usepackage{graphicx}
\usepackage{gensymb}
\usepackage{color}

\newcommand{\beq}{\begin{equation}}
\newcommand{\eeq}{\end{equation}}

\newcommand{\etal}{\textit{et al.}}
\newcommand{\mrm}{\mathrm}
\newcommand{\unit}{\,\mrm}

\newcommand{\kB}{k_\mrm{B}}
\newcommand{\Tc}{T_\mrm{c}}
\newcommand{\Pc}{P_\mrm{c}}

\newcommand{\taur}{\tau_\mathrm{r}}

\newcommand{\degreeC}{^\circ\mathrm{C}} 
\newcommand{\kb}{k_\mrm{B}}

\newcommand{\ELDS}{E_\mrm{LDS}}
\newcommand{\DvLDS}{\Delta v_\mrm{LDS}}
\newcommand{\TB}{T_0}
\newcommand{\EHDS}{E_\mrm{HDS}}
\newcommand{\DvHDS}{\Delta v_\mrm{HDS}}
\newcommand{\Tref}{T_\mrm{ref}}
\newcommand{\Tg}{T_\mrm{g}}
\newcommand{\Npts}{N_\mrm{points}}

\begin{document}
\title{Viscosity and self-diffusion of supercooled and stretched water from molecular dynamics simulations}

\author{Pablo Montero de Hijes}
\affiliation{Departamento de estructura de la materia, f\'{\i}sica t\'ermica y electr\'onica, Facultad de Ciencias Fisicas, Universidad Complutense de Madrid, 28040, Spain}
\author{Eduardo Sanz}
\affiliation{Departamento de estructura de la materia, f\'{\i}sica t\'ermica y electr\'onica, Facultad de Ciencias Fisicas, Universidad Complutense de Madrid, 28040, Spain}
\author{Laurent Joly}
\affiliation{Univ Lyon, Universit\'e Claude Bernard Lyon 1, CNRS, Institut Lumi\`{e}re Mati\`{e}re, F-69622, Lyon, France}
\author{Chantal Valeriani}
\affiliation{Departamento de estructura de la materia, f\'{\i}sica t\'ermica y electr\'onica, Facultad de Ciencias Fisicas, Universidad Complutense de Madrid, 28040, Spain}
\author{Fr\'ed\'eric Caupin}
\affiliation{Univ Lyon, Universit\'e Claude Bernard Lyon 1, CNRS, Institut Lumi\`{e}re Mati\`{e}re, F-69622, Lyon, France}

\date{\today}

\begin{abstract}
Among the numerous anomalies of water, the acceleration of dynamics under pressure is particularly puzzling. Whereas the diffusivity anomaly observed in experiments has been reproduced in several computer studies, the parallel viscosity anomaly has received less attention. Here we simulate viscosity and self-diffusion coefficient of the TIP4P/2005 water model over a broad temperature and pressure range. We reproduce the experimental behavior, and find additional anomalies at negative pressure. The anomalous effect of pressure on dynamic properties becomes more pronounced upon cooling, reaching two orders of magnitude for viscosity at $220\unit{K}$. We analyze our results with a dynamic extension of a thermodynamic two-state model, an approach which has proved successful in describing experimental data. Water is regarded as a mixture of interconverting species with contrasting dynamic behaviors, one being strong (Arrhenius), and the other fragile (non-Arrhenius). The dynamic parameters of the two-state models are remarkably close between experiment and simulations. The larger pressure range accessible to simulations suggests a modification of the dynamic two-state model, which in turn also improves the agreement with experimental data. Furthermore, our simulations demonstrate the decoupling between viscosity $\eta$ and self-diffusion coefficient $D$ as a function of temperature $T$. The Stokes-Einstein relation, which predicts a constant $D \eta/T$, is violated when $T$ is lowered, in connection with the Widom line defined by an equal fraction of the two interconverting species. These results provide a unifying picture of thermodynamics and dynamics in water, and call for experiments at negative pressure.
\end{abstract}

\maketitle

\section{Introduction\label{sec:intro}}

Liquid water exhibits countless thermodynamic and dynamic peculiarities~\cite{Gallo_water_2016}. Among thermodynamic properties, well known anomalies are the negative expansion coefficient below $4\degreeC$ at ambient pressure, or the rapid increase in isothermal compressibility and isobaric heat capacity upon cooling. These anomalies become more pronounced in supercooled water~\cite{Debenedetti_supercooled_2003,Holten_thermodynamics_2012}. Several dynamic properties are also anomalous, showing a non-monotonic pressure dependence. Below room temperature, the shear viscosity $\eta$ reaches a minimum as a function of pressure~\cite{Rontgen_ueber_1884,Warburg_ueber_1884,Bridgman_viscosity_1925,Bett_effect_1965}, whose location has been recently tracked down to $244\unit{K}$ and $200\unit{MPa}$, where $\eta$ is reduced by 42\% compared to its value at ambient pressure~\cite{Singh_pressure_2017}. Diffusivity reaches a maximum as a function of pressure, which has been measured in supercooled water both for translation~\cite{Prielmeier_pressure_1988,Harris_selfdiffusion_1997} and rotation~\cite{Lang_high_1981,Arnold_pressure_2002}. Stretched water, or water at negative pressure, has also been studied, although less extensively (see Ref.~\onlinecite{Caupin_escaping_2015} for a review). The temperature of maximum density increases from $4\degreeC$ at ambient pressure to $18\degreeC$ at $-137\unit{MPa}$, and a maximum in the isothermal compressibility of water along isobars has been revealed around $-100\unit{MPa}$ and below $276\unit{K}$~\cite{Holten_compressibility_2017}.

A limit to experiments on metastable water is homogeneous nucleation of ice in supercooled water, or of vapor in stretched water. At deeply metastable conditions, nucleation becomes unavoidable on the timescale needed to perform measurements. Because of the small sizes and short timescales involved, molecular dynamics (MD) simulations provide a powerful alternative to experiments for studying physical properties at even more extreme conditions. Extensive thermodynamic data are already available for several water models such as ST2~\cite{Poole_phase_1992,Poole_density_2005} and TIP4P/2005~\cite{Agarwal_thermodynamic_2011,Gonzalez_comprehensive_2016,Biddle_twostructure_2017}. The self-diffusion coefficient $D$ has also been studied in simulations. Early simulations reproduced qualitatively the experimental behavior of $D$: first its anomalous density dependence for ST2~\cite{Sciortino_effect_1991} and SPC/E water~\cite{Vaisman_mobility_1993}, and later its maximum for SPC/E water~\cite{Starr_dynamics_1999,Scala_configurational_2000,Errington_relationship_2001}. A minimum in $D$ at low density, not yet observed in experiments, has also been found in simulations of TIP4P~\cite{Ruocco_molecular_1993} and SPC/E water~\cite{Starr_dynamics_1999,Errington_relationship_2001,Netz_static_2001}. Agarwal~\etal~\cite{Agarwal_thermodynamic_2011} simulated $D$ of water for five models, namely SPC/E, mTIP3P, TIP4P, TIP5P, and TIP4P/2005. Although they all show a maximum in $D$ as a function of density at low enough temperature, only TIP4P/2005 gives a maximum at ambient temperature, as observed in experiments. All models give rise to a minimum in $D$ at low density. One concern about the results for $D$ is the possible existence of finite-size effects, with simulations involving for instance 256 molecules only~\cite{Agarwal_thermodynamic_2011}. Correcting for these effects requires the knowledge of the viscosity~\cite{Yeh_systemsize_2004,Tazi_diffusion_2012}.

However, simulations of viscosity are scarce. Because of its lower computational cost, the structural relaxation time $\tau_\alpha$ is often used as a proxy for $\eta$, as these two quantities are assumed to be proportional. However, Shi~\etal~\cite{Shi_relaxation_2013} found that, for model atomic and molecular systems, $\tau_\alpha/\eta$ is temperature dependent. The same issue was observed for a water model~\cite{Guillaud_decoupling_2017,Guillaud_assessment_2017}. Coming back to direct simulations of $\eta$, we list here the important works relevant to our study. A minimum in the density dependence of $\eta$ was obtained with TIP4P/2005~\cite{Gonzalez_shear_2010} and BK3 water~\cite{Kiss_anomalous_2014}. Values of $D$ and $\eta$ for TIP4P/2005 were also reported~\cite{Guevara-Carrion_prediction_2011} in the range $260$--$400\unit{K}$ and $0.1$--$300\unit{MPa}$, and showed the maximum in $D$, whereas the minimum in $\eta$ was hidden by the simulation uncertainties. To our knowledge, simulation data for $\eta$ of TIP4P/2005 water at supercooled conditions are only available at ambient pressure~\cite{Guillaud_decoupling_2017} or a density of $1000\unit{kg\,m^{-3}}$.~\cite{Kawasaki_identifying_2017} We are aware of only two simulation studies of viscosity in the supercooled region under pressure. The first by Dhabal~\etal~\cite{Dhabal_comparison_2016} reported $D$ and $\eta$ for the coarse-grained mW model (monatomic water), and the density dependence gave a minimum and a maximum for $D$ and a minimum for $\eta$. However, because it omits the reorientation of hydrogen atoms, mW gives $D$ three times higher and $\eta$ three times lower than experimental values for water at ambient conditions. The second study simulated the more realistic WAIL potential~\cite{Ma_continuous_2015}, but the pressure range investigated ($0$--$70\unit{MPa}$) precluded the observation of a minimum in $\eta$.

It is therefore of interest to perform simulations with a realistic water model, aimed at the direct determination of $\eta$ in a broad pressure and temperature range. In particular, it should be possible to follow in the supercooled region the minimum in $\eta$ seen at stable conditions, and also to investigate if there is a maximum in $\eta$ at low density, similar to the second extremum seen in simulations of $D$. In the present work, we have performed such simulations with TIP4P/2005 water. We have computed $\eta$ and $D$ at the same state points, so that we were able at the same time to apply finite-size corrections to $D$.

An additional motivation of our work is to investigate the connection between thermodynamics and dynamics. In the case of real water, several works have addressed this question using a two-state model for theoretical frame~\cite{Vedamuthu_properties_1994,Cho_molecularlevel_1999,Cho_pressure_2002,Tanaka_simple_2000,Tanaka_new_2003}. In Ref.~\onlinecite{Singh_pressure_2017}, an accurate thermodynamic two-state model~\cite{Holten_equation_2014} was successfully extended to describe dynamic data. As a similar thermodynamic two-state model is available for TIP4P/2005 water~\cite{Biddle_twostructure_2017}, we investigate here if its dynamic extension can also reproduce our simulated dynamic properties.

Finally, obtaining simultaneous data on $D$ and $\eta$ is also useful to test their coupling. Indeed, in liquids at high temperature, $D$ and $\eta$ are usually linked by the Stokes-Einstein (SE) relation, inspired by macroscopic hydrodynamics and linear response theory, which states that $D \eta / T$ is independent of temperature. Deviations are observed in supercooled liquids, usually around 1.3 $\Tg$ where $\Tg$ is the glass transition temperature; see for instance Ref.~\onlinecite{Chang_heterogeneity_1997} for $D \eta / T$ vs. $T$ for six glassformers. In contrast, at ambient pressure, water already exhibits a violation of the SE relation at room temperature (above $2\Tg$); this violation increases upon cooling, with a relative deviation around 70\% at $239\unit{K}$ \cite{Dehaoui_viscosity_2015}. Understanding the origin of this early SE violation in water is an active field of research~\cite{Guillaud_decoupling_2017,Galamba_hydrogenbond_2017,Kawasaki_identifying_2017}, as for other anomalies of water that become more pronounced in the supercooled region~\cite{Gallo_water_2016}.

\section{Methods\label{sec:methods}}

\begin{table*}[tttt]
\centering
\caption{\label{tab:bestfit1} Best fit parameters of the original two-state model for dynamic properties (Eq.~\ref{eq:fit1}), applied to simulation set 1 (this work) and to the experiment~\cite{Singh_pressure_2017}. A common temperature $\TB$ is used for the different dynamic properties. Uncertainties correspond to a 95\% confidence interval. The number of points and reduced $\chi^2$ are also given.}
\begin{tabular}{ccccccccccc}
\hline
\hline
		&	\multicolumn{2}{c}{Simulations} &	\multicolumn{3}{c}{Experiment}	\\
Quantity & Viscosity	& Self-diffusion	& Viscosity	& Self-diffusion	& Rotational\\
& $\eta$ & coefficient $D$	& $\eta$ & coefficient $D$	& correlation time $\taur$\\
\hline
$A_0$	& $37.19\pm 1.32\unit{\mu Pa\,s}$ & $39715 \pm 950\unit{\mu m^2\,s^{-1}}$	& $38.75\pm 0.63\unit{\mu Pa\,s}$ & $40330 \pm 320\unit{\mu m^2\,s^{-1}}$	& $86.2 \pm 3.0\unit{fs}$	\\
$\ELDS/\kb (\mrm{K})$	& $1874 \pm 56$	& $2034 \pm 21$	& $2262 \pm 23$	& $1984 \pm 21$	& $2585 \pm 53$	\\
$\EHDS/\kb (\mrm{K})$	& $350.2 \pm 10.2$	& $288.0 \pm 5.1$	& $421.9 \pm 3.2$	& $402.2 \pm 1.5$	& $395.0 \pm 5.5$	\\
$\DvHDS (10^{-30}\unit{m^3})$	& $3.32 \pm 0.25$	& $3.80 \pm	0.11$	& $2.44 \pm 0.08$	& $1.79 \pm	0.04$	& $1.62 \pm 0.13$	\\
$\TB (\unit{K})$	&\multicolumn{2}{c}{$145.86$} &	\multicolumn{3}{c}{$147.75$}	\\
$\Npts$	& $26$	& $26$	& $178$	& $157$	& $101$\\
$\chi^2$	& $3.30$	& $7.54$	& $1.57$	& $1.48$	& $0.82$\\
\hline
\hline
\end{tabular}
\end{table*}

\subsection{Simulation details\label{sec:simdetails}}

We have selected the TIP4P/2005 model for water~\cite{Abascal_general_2005}, which is currently one of the best force fields available, describing nearly quantitatively many properties of water in a broad temperature and pressure range. Many thermodymanic quantities are available for TIP4P/2005 water, and they have been successfully described within the two-state formalism by Biddle \etal~\cite{Biddle_twostructure_2017} (see Section~\ref{sec:2statesmodel}). We have performed $NVT$ runs of TIP4P/2005 water simulated via the LAMMPS MD package~\cite{Plimpton_fast_1995}. $N$ is set to 216 molecules and the temperature is kept constant via a Nos\'{e}-Hoover thermostat. To remain consistent with the definition of TIP4P/2005~\cite{Abascal_general_2005}, we used a $0.85\unit{nm}$ cutoff. Long-range Coulombic interactions were computed using the particle-particle particle-mesh method~\cite{Hockney_computer_1988}, and water
molecules were held rigid using the SHAKE algorithm~\cite{Ryckaert_numerical_1977}. We simulate temperatures ranging from $220$ to $300\unit{K}$ and densities from $800$ to $1320\unit{kg\,m^{-3}}$. We selected state points on a grid in the temperature-density plane, which includes the validity region of the thermodynamic two-state model by Biddle \etal~\cite{Biddle_twostructure_2017}. All state points have been simulated far beyond their characteristic time to ensure equilibration (see for instance Fig. 1B of Ref.~\citenum{Kawasaki_identifying_2017} for characteristic times of TIP4P/2005 water at $1000\unit{kg\,m^{-3}}$). The run durations range from $25\unit{ns}$ at $1320\unit{kg\,m^{-3}}$ and $300\unit{K}$ to $88\unit{ns}$ at $960\unit{kg\,m^{-3}}$ and $220\unit{K}$; at $920\unit{kg\,m^{-3}}$ and $220\unit{K}$, a longer duration of $880\unit{ns}$ was used. For each state point, we obtain the shear viscosity $\eta$ by averaging the five independent Green-Kubo integrals of the auto-correlation function of traceless stress tensor elements~\cite{Chen_are_2009}. As these calculations were computationally expensive, optimized algorithms were used~\cite{Ramirez_efficient_2010}. We calculate the self-diffusion coefficient $D$ from the slope of the linear regression of the mean squared displacement $\langle r^2 \rangle$ in the diffusive regime. The slope is divided by 6 following the Einstein relation~\cite{Allen_computer_2017}:  $\langle r^2 \rangle = 6Dt$ to obtain $D$ (note that center of mass corrections have been used). Because of hydrodynamic interactions between image boxes in a simulation with periodic boundary conditions, the raw value of $D$ suffers from finite size effects. It has been shown on theoretical grounds and verified with simulations of boxes with different sizes~\cite{Yeh_systemsize_2004,Tazi_diffusion_2012}, that the value for the self-diffusion in an infinite liquid can be calculated with the following formula:
\beq
D= D_\mrm{PBC} + 2.837 \frac{\kB T}{6 \pi \eta L_\mrm{box}} \; ,
\label{eq:Dcorr}
\eeq
where $D_\mrm{PBC}$ is the self-diffusion coefficient before finite size correction (that is in a cubic simulation box of side $L_\mrm{box}$ with periodic boundary conditions), $\kB$ the Boltzmann constant, $T$ the temperature, and $\eta$ the viscosity (previously obtained from the simulation). Tazi~\etal~\cite{Tazi_diffusion_2012} also simulated TIP4P/2005 water but for only one state point. They computed $D_\mrm{PBC}$ for several box sizes $L_\mrm{box}$, and used Eq.~\ref{eq:Dcorr} to calculate $D$ for the infinite liquid. From the slope of $D_\mrm{PBC}$ vs. $1/L_\mrm{box}$, they also obtained an estimate of $\eta$, which was in perfect agreement with $\eta$ directly calculated from the Green-Kubo integrals. This validates our procedure of first calculating $\eta$ from the Green-Kubo integrals and $D_\mrm{PBC}$ for one value of $L_\mrm{box}$ (e.g. $1.863\unit{nm}$ for $\rho=1000\unit{kg\,m^{-3}}$), and then using $\eta$ and Eq.~\ref{eq:Dcorr} to calculate $D$ for the infinite liquid. Appendix~\ref{sec:data} gives all simulations results for $\eta$ (Table~\ref{tab:dataeta}) and for $D$ (Table~\ref{tab:dataD}). We also present in Appendix~\ref{sec:data} how uncertainties on $\eta$ and $D$ were estimated; their values are given in the tables.

\begin{figure}[ttt]
\centering
\centerline{\includegraphics[width=0.9\columnwidth]{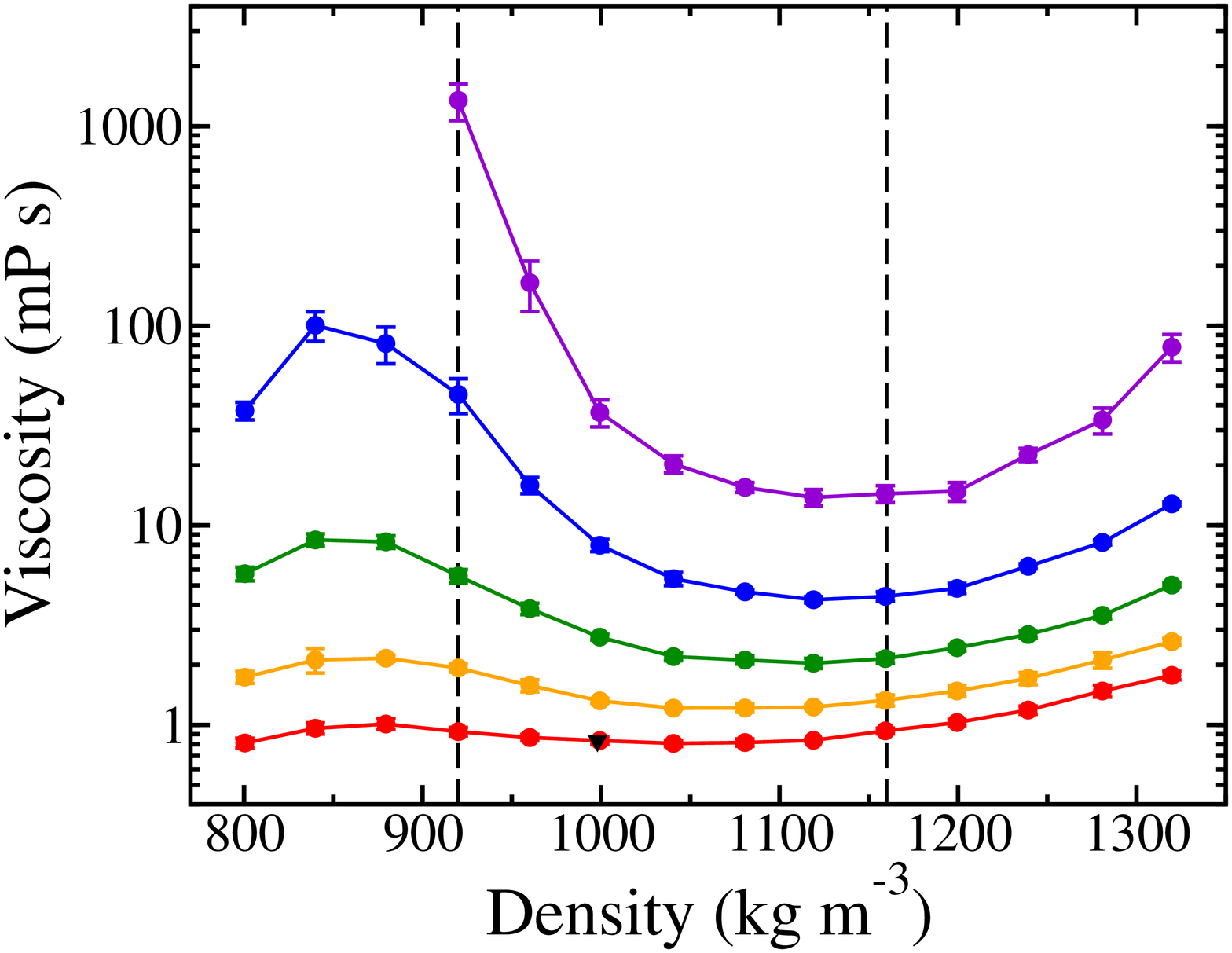}}
\centerline{\includegraphics[width=0.9\columnwidth]{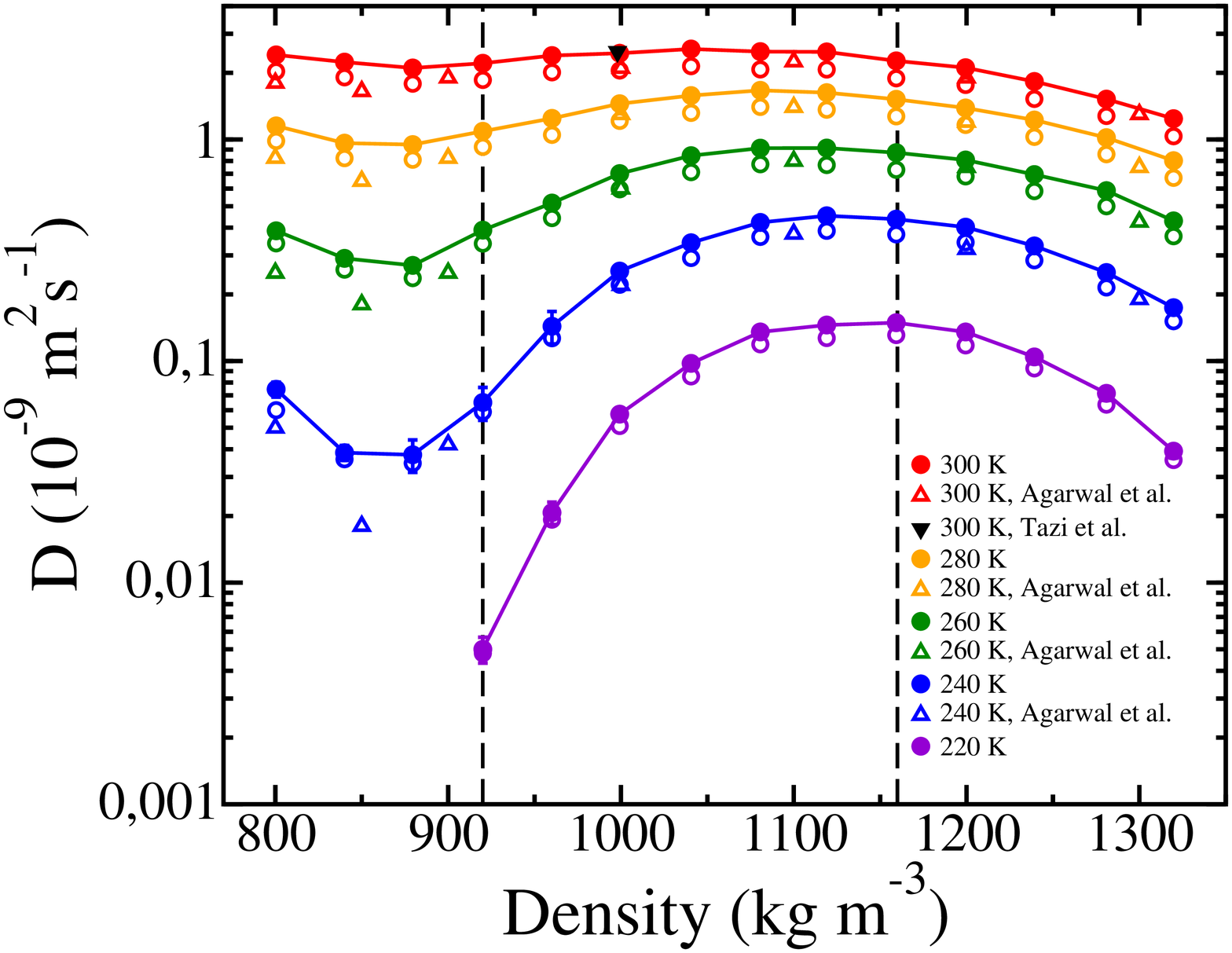}}
\caption{\textbf{Density dependence of viscosity (a) and self-diffusion coefficient (b) along several isotherms}. The data set for each isotherm (circles: this work; down triangle: Ref.~\onlinecite{Tazi_diffusion_2012}; up triangles: Ref.~\onlinecite{Agarwal_thermodynamic_2011}) is shown with a distinct color and labeled with the temperature in K. In (b), the empty and filled symbols correspond to data before and after correction with Eq.~\ref{eq:Dcorr}, respectively. The solid lines connecting points are guides to the eye. The dashed lines bracket the validity region of the two-state model from Ref.~\onlinecite{Biddle_twostructure_2017}.\label{fig:eta-D_all}}
\end{figure}

\begin{figure}[ttt]
\centering
\centerline{\includegraphics[width=0.85\columnwidth]{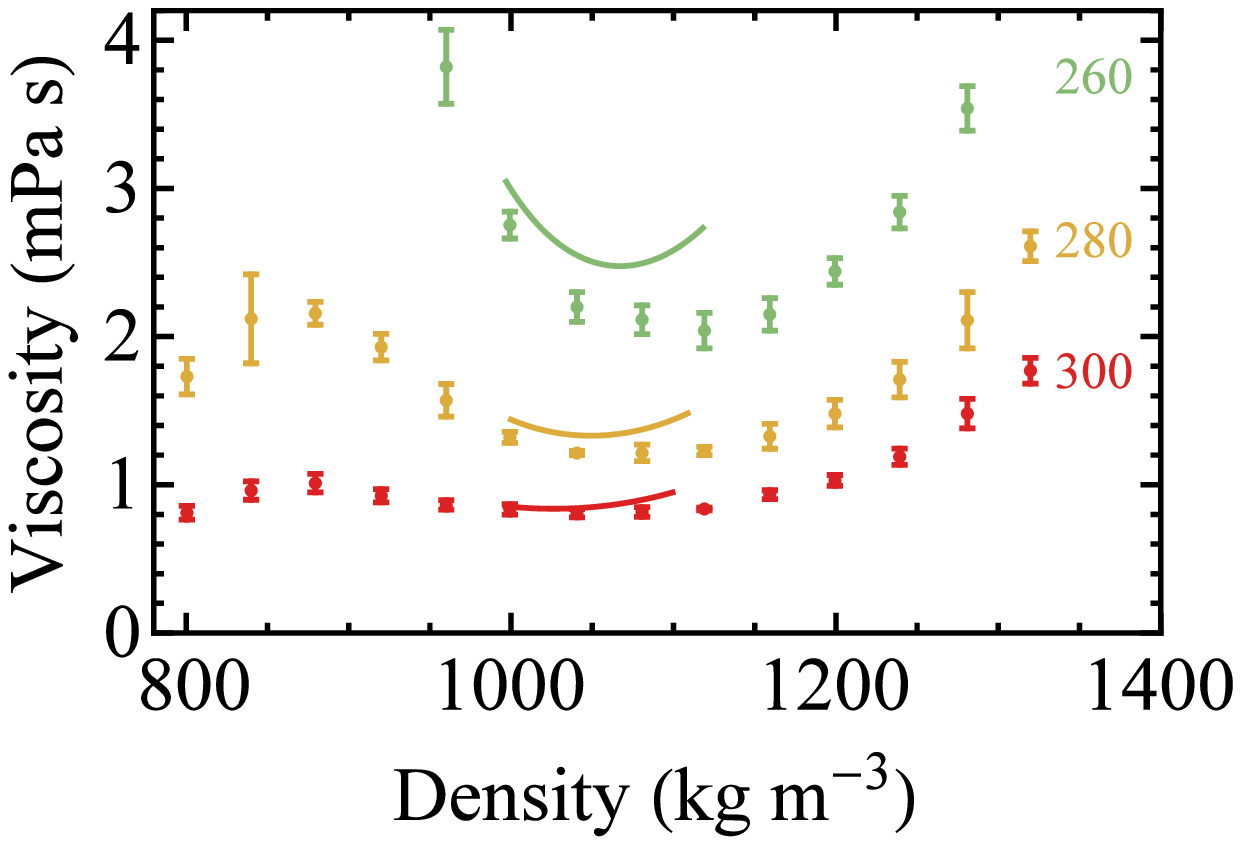}}
\centerline{\includegraphics[width=0.85\columnwidth]{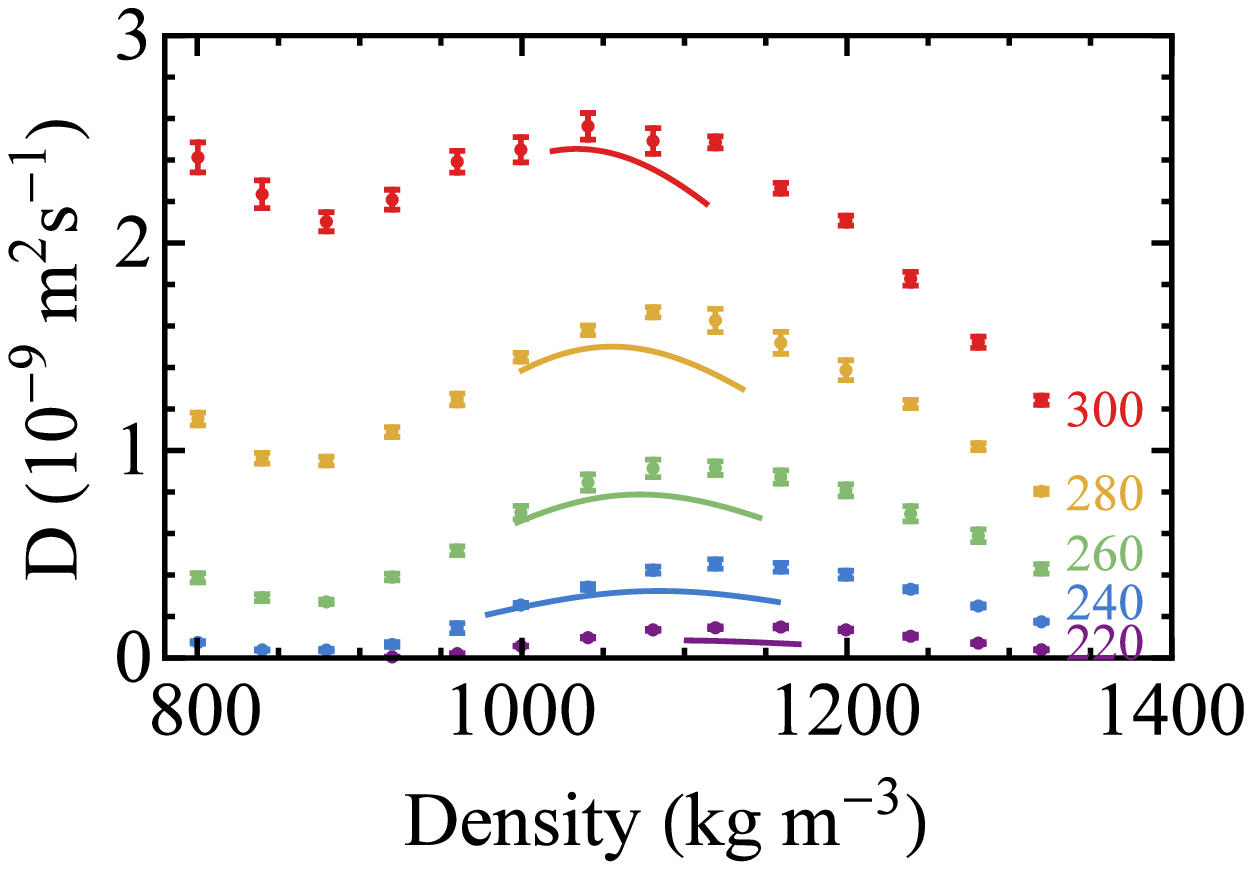}}
\caption{Details of previous figure showing our simulations results (circles) on a linear vertical scale, and experimental data~\cite{Singh_pressure_2017} (solid lines, see text for details).\label{fig:eta-D_zoom}}
\end{figure}

\subsection{Two-state model\label{sec:2statesmodel}}

Two-state models are popular explanations of the anomalies of water, because anomalous behavior in such models stems from the variation of the fraction of each state, each having otherwise a normal behavior. For instance, Robinson and his colleagues provided a two-state description of density at ambient pressure~\cite{Vedamuthu_properties_1994}, later extended to the pressure dependence of viscosity~\cite{Cho_molecularlevel_1999} and density~\cite{Cho_pressure_2002}. A more comprehensive description was formulated by Tanaka~\cite{Tanaka_simple_2000,Tanaka_new_2003} to account for the anomalous behavior of density, isothermal compressibility, isobaric heat capacity, and shear viscosity with a mixture of two states with fractions $f(T,P)$ and $1-f(T,P)$. The dynamic part of Tanaka's model describes the viscosity of water as a thermally activated process, whose activation energy $E_\mrm{a}$ is the fraction-weighted average of the activation energy for each state, $E_1$ and $E_2$: $E_\mrm{a}= f(T,P) E_1 + [1-f(T,P)] E_2$. In other words, the hypothetic liquids made of each pure state would have an Arrhenius behavior (constant $E_1$ and $E_2$), and the non-Arrhenius behavior of real water would arise from the variation of the fraction $f(T,P)$. Holten, Sengers and Anisimov~\cite{Holten_equation_2014} developed an equation of state for water based on the two-state picture (HSA model). In the HSA model, water is considered as an athermal non-ideal `solution' of two rapidly inter-convertible states or structures: a low density state (LDS) and a high density state (HDS), with respective fractions $f$ and $1-f$. The non-ideality of the solution drives a first-order phase transition between two distinct liquids at low temperature, ending at a liquid-liquid critical point (LLCP) at $\Tc=228.2\unit{K}$ and $\Pc=0\unit{MPa}$. We emphasize that there is currently no firmly established experimental proof of such a liquid-liquid transition and LLCP for real water, the main reason being that, in experiments, ice nucleates before reaching the putative two-phase region~\cite{Caupin_escaping_2015}. Nevertheless, the HSA model achieves a fit within experimental error of a comprehensive data set of thermodynamic properties (density, isothermal compressibility, thermal expansion coefficient, isobaric heat capacity, and speed of sound) in the range 200 to $310\unit{K}$ and 0.1 to $400\unit{MPa}$. This equation of state is the current official guideline on thermodynamic properties of supercooled water~\cite{TheInternationalAssociationforthePropertiesofWaterandSteam_Guideline_2015}. Following Tanaka's example~\cite{Tanaka_simple_2000,Tanaka_new_2003}, the HSA model was recently extended to dynamic properties by Singh \etal~\cite{Singh_pressure_2017}, who additionally measured viscosity of supercooled water under pressure. Experimental data for stable and supercooled water below $300\unit{K}$ and between $0$ and $400\unit{MPa}$ were included, not only for shear viscosity as Tanaka did~\cite{Tanaka_simple_2000,Tanaka_new_2003}, but also for self-diffusion coefficient\cite{Prielmeier_pressure_1988,Harris_selfdiffusion_1997} and rotational correlation time~\cite{Lang_high_1981,Arnold_pressure_2002}. It was observed that a mixture of two liquids following Arrhenius dynamics did not give satisfactory results. Instead, all properties could be reproduced within experimental uncertainty if the high density state was assumed to follow a fragile behavior. Eventually the following form was used to describe all three dynamic properties:
\begin{eqnarray}
&A(T,P) = A_0 \left( \frac{T}{\Tref} \right)^\nu \nonumber \\
&\exp \left\{\epsilon \left[ \left[1 -f(T,P) \right] \frac{\EHDS + \DvHDS P}{\kb (T- \TB)} + f(T,P) \frac{\ELDS}{\kb T} \right]\right\} \; .
\label{eq:fit1}
\end{eqnarray}
Here $\Tref = 273.15\unit{K}$ (introduced to make $T/\Tref$ dimensionless), $\nu$ accounts for the temperature variation of the average speed of the molecules~\cite{Prielmeier_pressure_1988} ($\nu=1/2$ for $A=\eta$ or $D$, $-1/2$ for $A=\taur$\footnote{The value of $\nu$ for $\taur$ is chosen for consistency with the Stokes-Einstein-Debye relation $T \taur/\eta = \mrm{cst}_1$, which holds at high temperature, similar to the Stokes-Einstein relation $D\eta / T= \mrm{cst}_2$~\cite{Dehaoui_viscosity_2015}.}), and $\epsilon=1$ for $A=\eta$ or $\taur$ and $-1$ for $A=D$. There are also 5 free parameters, as for Tanaka's viscosity model. Their physical meaning is as follows. $A_0$ is a global scale factor. LDS behaves like an Arrhenian liquid with activation energy $\ELDS$, whereas HDS behaves like a fragile liquid described by a Vogel-Tamann-Fulcher (VTF) law with parameters $\EHDS + \DvHDS P$ and $\TB$. The energy appearing in the VTF law has a pressure dependence $\DvHDS P$ coming from the difference in volume between the activated and initial state of the activated process~\cite{Tanaka_simple_2000}. A good fit, with good reduced $\chi^2$ and residuals, could be obtained holding $\TB$ equal for the three properties (see Fig.~3 of Ref.~\onlinecite{Singh_pressure_2017}). The best fit parameters are reproduced in Table~\ref{tab:bestfit1}. They are relatively close between properties, and have reasonable physical values. In particular, $\ELDS$ is of the order of the hydrogen bond energy, and $\DvHDS$ is around 5-8\% of the volume per molecule, around $30\,10^{-30}\unit{m^3}$ at $\rho=1000\unit{kg\,m^{-3}}$.

One focus of the present paper is applying a two-state approach for simulation data. Recently, a set of thermodynamic properties of TIP4P/2005 water was successfully described with a two-state model similar to the HSA model~\cite{Biddle_twostructure_2017}. Its validity region (Fig.~1 of Ref.~\onlinecite{Biddle_twostructure_2017}) covers temperatures from $180$ to $320\unit{K}$ and pressures from around $-250$ to $500\unit{MPa}$. It predicts a liquid-liquid critical point at $182\unit{K}$ and $170\unit{MPa}$. These values are close to previous estimates for TIP4P/2005~\cite{Abascal_widom_2010,Sumi_effects_2013,Yagasaki_spontaneous_2014,Singh_twostate_2016}. Although the existence of such a critical point in TIP4P/2005 has been challenged~\cite{Overduin_fluctuations_2015}, a recent approach based on potential energy landscape~\cite{Handle_potential_2018} also predicts a critical point. Inspired by the analysis performed on experimental data~\cite{Singh_pressure_2017}, we have investigated if, for simulations, the two-state model presented in Ref.~\onlinecite{Biddle_twostructure_2017} could be extended to describe dynamic properties.

\begin{figure*}[ttt]
\centerline{
\includegraphics[height=11.2cm]{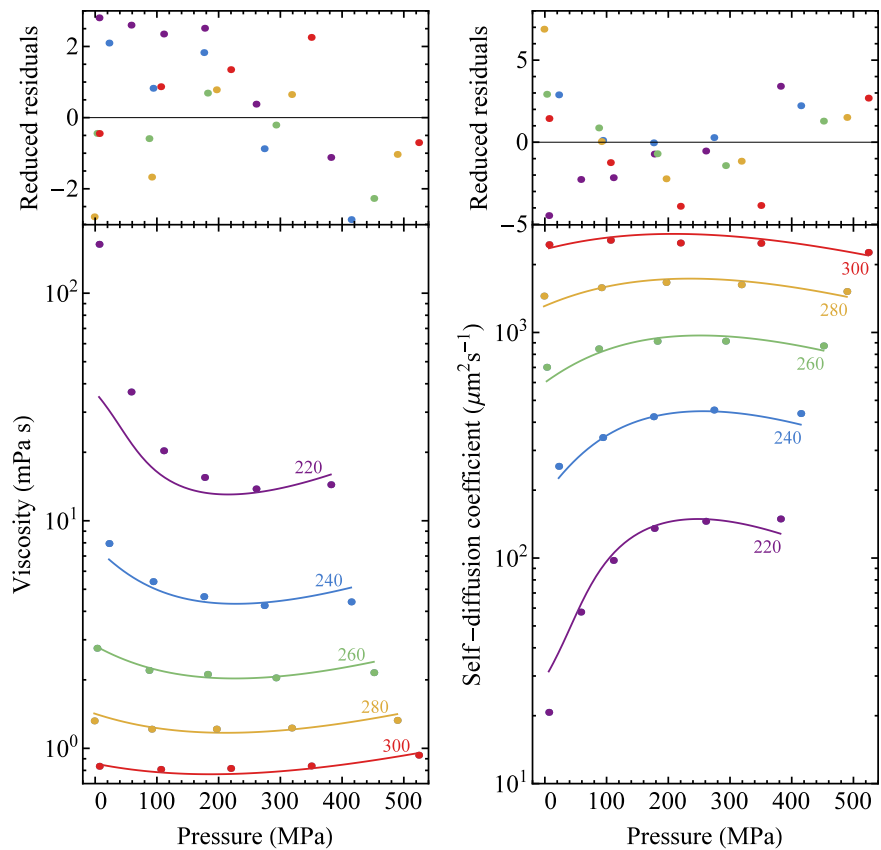}
}
\caption{\textbf{Pressure dependence of simulated dynamic properties (set 1) and original dynamic two-state model (Eq.~\ref{eq:fit1})}. The best fits to Eq.~\ref{eq:fit1} for the simulation set 1 are shown for viscosity (left) and self-diffusion coefficient (right). Best fit parameters are given in Table~\ref{tab:bestfit1}. In the bottom panels, the differently colored curves labeled by the temperature in K correspond to the values calculated along isotherms. The top panels show the deviations between fitted values and data points, each normalized by the simulation uncertainty (one standard deviation). \label{fig:simulfit1}}
\end{figure*}

\begin{figure*}[ttt]
\centerline{
\includegraphics[height=11.2cm]{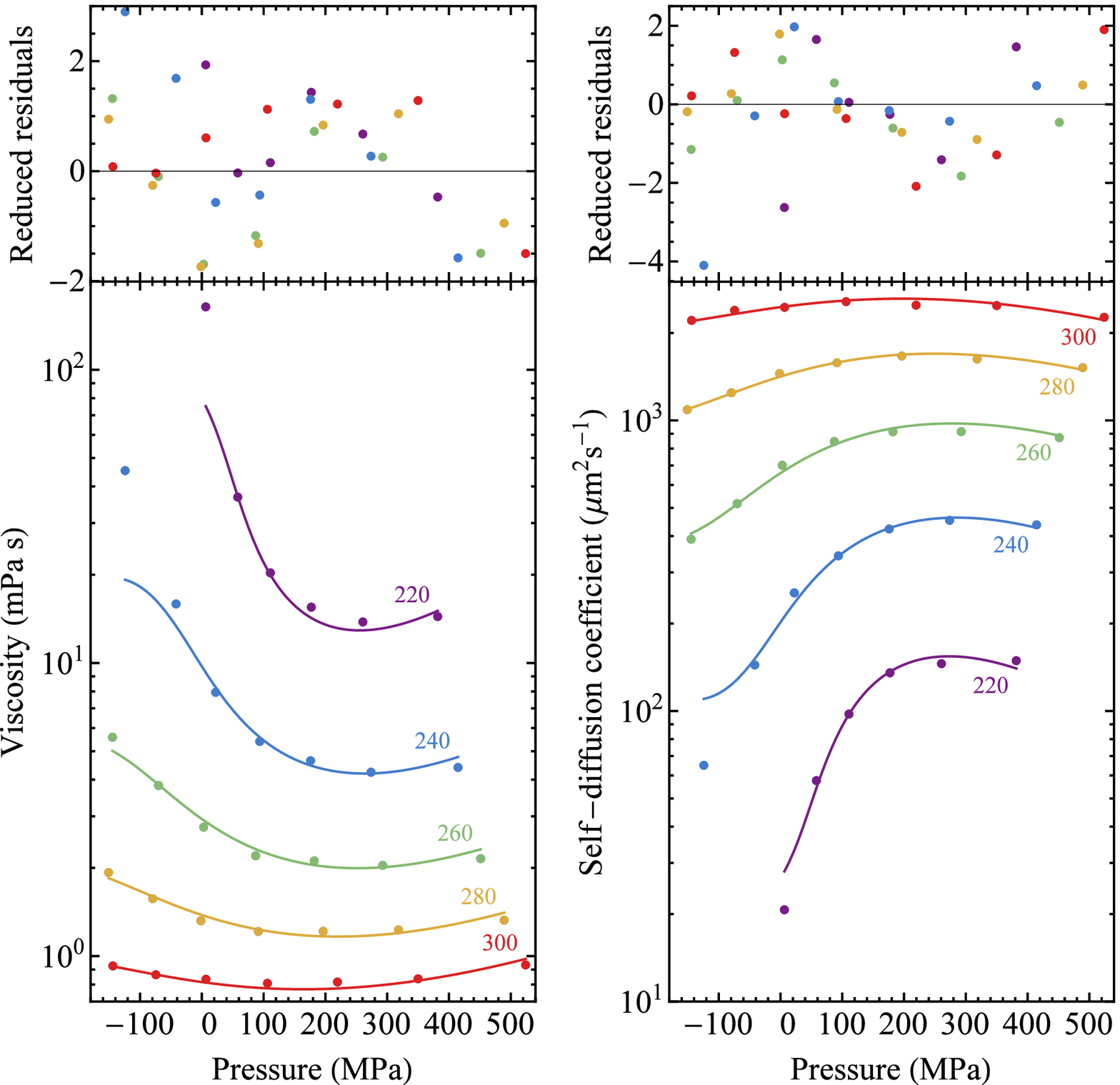}
}
\caption{\textbf{Pressure dependence of simulated dynamic properties (set 2) and modified dynamic two-state model (Eq.~\ref{eq:fit2})}. Same as Fig.~\ref{fig:simulfit1} for the fitting to Eq.~\ref{eq:fit2} of the simulation set 2. Best fit parameters are given in Table~\ref{tab:bestfit2}.\label{fig:simulfit2}}
\end{figure*}

\section{Results and discussion\label{sec:results}}

\subsection{Simulation results\label{sec:simresults}}

Figure~\ref{fig:eta-D_all} shows the final results for $\eta$ and $D$ as a function of density for a series of isotherms. Our results compare well with those of Tazi~\etal~\cite{Tazi_diffusion_2012} for both $\eta$ and $D$. Our uncorrected values for $D$ agree well with Agarwal~\etal~\cite{Agarwal_thermodynamic_2011} at high density. A slight discrepancy appears at low density and gets more pronounced at low temperature. Note that the difference with Ref.~\cite{Agarwal_thermodynamic_2011} is that we could correct $D$ for finite size effects because we have both $\eta$ and $D$. Figure~\ref{fig:eta-D_zoom} shows a close-up, to allow comparison with experimental data. The fits of Ref.~\onlinecite{Singh_pressure_2017} were used to represent the experimental data. Simulations reproduce well the fast temperature variation of $\eta$ and $D$, together with their minimum and maximum as a function of density, respectively. This illustrates once more the good performance of the TIP4P/2005 model in reproducing the properties of experimental water. At lower densities, where no experiment are available at present, our simulations yield a maximum in $\eta$ versus $\rho$ and a minimum in $D$ versus $\rho$. The minimum in $D$ has been previously observed in simulations~\cite{Ruocco_molecular_1993,Starr_dynamics_1999,Errington_relationship_2001,Netz_static_2001,Agarwal_thermodynamic_2011,Dhabal_comparison_2016}. To our knowledge, the maximum in $\eta$ is found here for the first time. The anomalous density variation (decrease of $\eta$ and increase of $D$) at fixed temperature becomes more pronounced upon cooling, as observed in the experiment (Fig.~\ref{fig:eta-D_zoom}). The anomalous change measured experimentally corresponds to a maximum factor $1.7$ for $\eta$ at $244\unit{K}$~\cite{Singh_pressure_2017} and $1.8$ for $D$ at $238\unit{K}$~\cite{Prielmeier_pressure_1988}. Because the simulations reach lower temperatures and densities, the observed factors reach larger values. At $220\unit{K}$, the anomalous change corresponds to a factor $98$ for $\eta$ and $30$ for $D$; note that these values are lower bounds, as no low density extremum is present in the density range of our simulations at this temperature.

To illustrate the fragile character of TIP4P/2005 water, Appendix~\ref{sec:Arrhenius} shows the variation of $\eta$ and $D$ with inverse temperature in a log-lin plot for each isochore (Arrhenius plots). Arrhenius behavior would correspond to straight lines. Instead, the curves exhibit a more rapid variation with decreasing temperature. The effect tends to be more pronounced at lower densities.

\subsection{Two-state analysis\label{sec:2statesresults}}


The analysis of the simulation data with the two-state model~\cite{Biddle_twostructure_2017} presented in Section~\ref{sec:2statesmodel} can be done only for state points in the validity region of the two-state model (between dashed vertical lines in Fig.~\ref{fig:eta-D_all}). Therefore, only data with density between $920$ and $1160\unit{kg\,m^{-3}}$ were considered. Because the dynamic two-state model (Eq.~\ref{eq:fit1}) uses pressure as a variable, the pressure for each state point was calculated from its temperature and density, using the thermodynamic two-state model~\cite{Biddle_twostructure_2017}. As a first step, we have tried to reproduce the analysis of experimental data (see Section~\ref{sec:2statesmodel}). To this end, we have selected a subset of simulation data, set 1, at positive pressure as in the experiment. Because its pressure was very close to zero, we also included in set 1 a data point at $280\unit{K}$ and $-1.5\unit{MPa}$. The fit to Eq.~\ref{eq:fit1} and the corresponding residuals are shown in Fig.~\ref{fig:simulfit1}, which corresponds to the simulation equivalent of Fig.~3 of Ref.~\onlinecite{Singh_pressure_2017} for the experiments. Overall the fit quality is reasonable. The reduced residuals, defined as the difference between data and fit values divided by the data uncertainty, are acceptable, but a systematic deviation appears at low temperature and low density. Table~\ref{tab:bestfit1} gives the best fit parameters. It can be seen that, as noted in Ref.~\onlinecite{Singh_pressure_2017} for the experiment, and here as well for the simulation set 1, the values of $\ELDS$, $\EHDS$, and $\DvHDS$ are in the same range for the different dynamic quantities. Note that they cannot have a common value for all dynamic properties, otherwise the SE relation would always hold. Moreover, the best fit parameters for the same dynamic quantity have similar values in simulations and in experiment. This confirms the good performance of the TIP4P/2005 model in reproducing the properties of experimental water. Remarkably, both in simulations and in experiment, the temperature $\TB$ is around $147\unit{K}$, and $\ELDS/\kB$ is in the range $1800$--$2300\unit{K}$, the typical energy of a hydrogen bond. The activation volume $\DvHDS$ is in the range $1.6$--$3.8\,10^{-30}\unit{m^3}$. This is around $5$--$12$\% of the volume per molecule in the liquid, around $30\,10^{-30}\unit{m^3}$ at $\rho=1000\unit{kg\,m^{-3}}$.

\begin{figure*}[ttt]
\centerline{
\includegraphics[height=11.2cm]{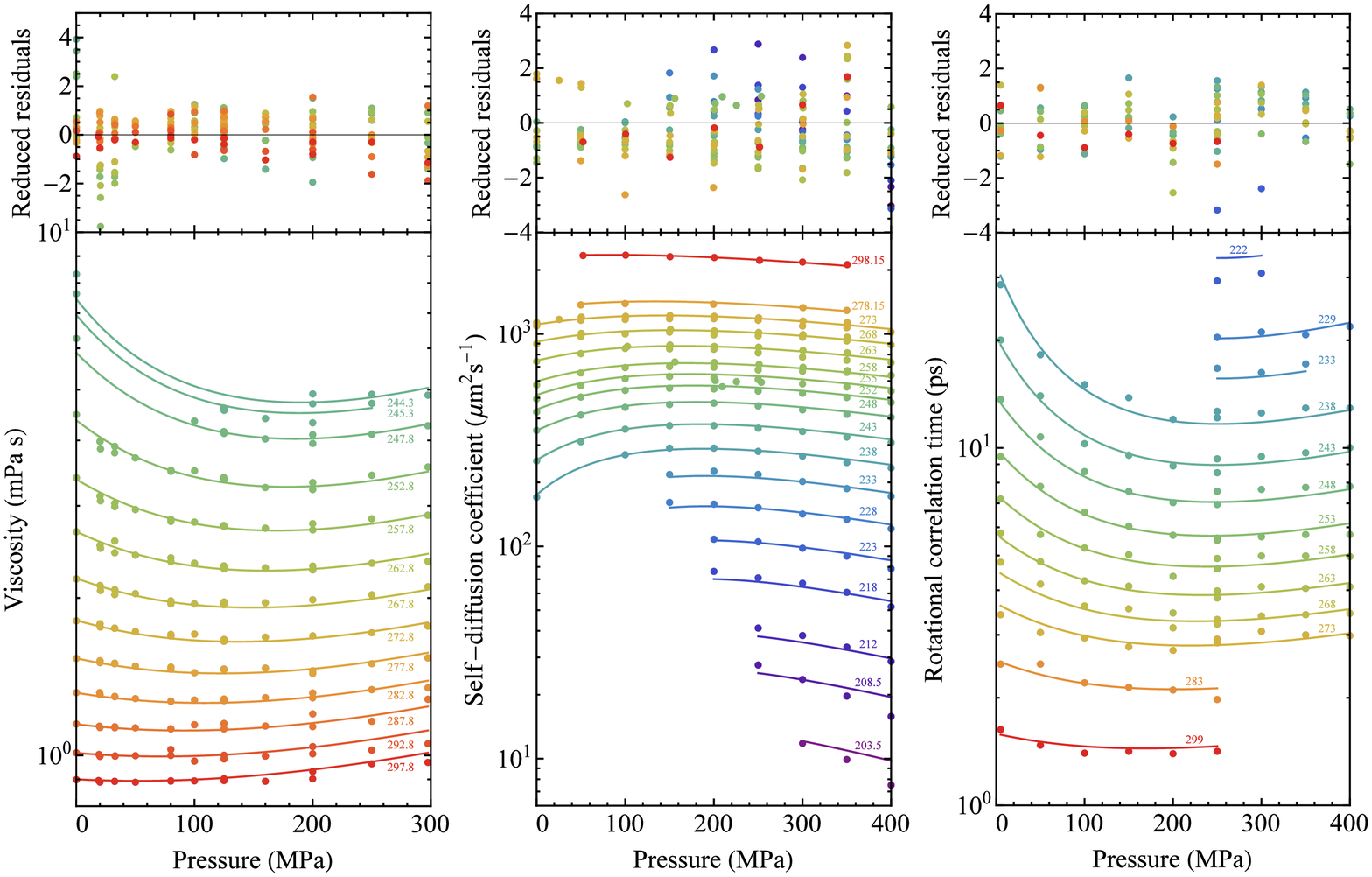}
}
\caption{\textbf{Pressure dependence of experimental dynamic properties and modified two-state model (Eq.~\ref{eq:fit2})}. The best fits to Eq.~\ref{eq:fit2} for the experimental data are shown for viscosity (left), self-diffusion coefficient (center), and rotational correlation time (right). Best fit parameters are given in Table~\ref{tab:bestfit2}. In the bottom panels, the differently colored curves labeled by the temperature in K correspond to the values calculated along isotherms. The top panels show the deviations between fitted values and data points, each normalized by the simulation uncertainty (one standard deviation).\label{fig:expfit2}}
\end{figure*}

\begin{table*}[tttt]
\centering
\caption{\label{tab:bestfit2} Best fit parameters of the modified two-state model for dynamic properties (Eq.~\ref{eq:fit2}), applied to simulation set 2 (this work) and to the experiment~\cite{Singh_pressure_2017}. A common temperature $\TB$ is used for the different dynamic properties. Uncertainties correspond to a 95\% confidence interval. The number of points and reduced $\chi^2$ are also given.}
\begin{tabular}{ccccccccccc}
\hline
\hline
		&	\multicolumn{2}{c}{Simulations} &	\multicolumn{3}{c}{Experiment}	\\
Quantity & Viscosity	& Self-diffusion	& Viscosity	& Self-diffusion	& Rotational\\
& $\eta$ & coefficient $D$	& $\eta$ & coefficient $D$	& correlation time $\taur$\\
\hline
$A_0$	& $60.23\pm 2.02\unit{\mu Pa\,s}$ & $24315 \pm 530\unit{\mu m^2\,s^{-1}}$	& $48.79\pm 1.16\unit{\mu Pa\,s}$ & $37280 \pm 350\unit{\mu m^2\,s^{-1}}$	& $93.3 \pm 3.8\unit{fs}$	\\
$\ELDS/\kb (\mrm{K})$	& $2239 \pm 53$	& $2067 \pm 22$	& $2433 \pm 28$	& $2056 \pm 30$	& $2626 \pm 71$	\\
$\DvLDS (10^{-30}\unit{m^3})$	& $28.9 \pm 2.8$	& $28.5 \pm	1.3$	& $42.5 \pm 4.0$	& $16.6 \pm	4.3$	& $15.1 \pm 14.8$	\\
$\EHDS/\kb (\mrm{K})$	& $182.0 \pm 10.5$	& $164.0 \pm 5.2$	& $376.3 \pm 4.5$	& $382.0 \pm 2.6$	& $375.7 \pm 9.3$	\\
$\DvHDS (10^{-30}\unit{m^3})$	& $4.29 \pm 0.19$	& $3.7 \pm	0.09$	& $2.69 \pm 0.08$	& $1.94 \pm	0.06$	& $1.76 \pm 0.20$	\\
$\TB (\unit{K})$	&\multicolumn{2}{c}{$158.55$} &	\multicolumn{3}{c}{$149.18$}	\\
$\Npts$	& $34$	& $34$	& $178$	& $157$	& $101$\\
$\chi^2$	& $1.61$	& $2.01$	& $0.94$	& $1.40$	& $0.85$\\
\hline
\hline
\end{tabular}
\end{table*}

As a second step, we attempted to fit all simulation data belonging to the validity region of the two-state model~\cite{Biddle_twostructure_2017}. The fit to Eq.~\ref{eq:fit1} deteriorates gradually when simulation data with lower density are successively added. Eq.~\ref{eq:fit1} cannot generate a low-density extremum in dynamic quantities. Fig.~\ref{fig:eta-D_all} shows that these extrema lie outside the region of validity of the two-state model~\cite{Biddle_twostructure_2017}, but still their vicinity might be responsible for the discrepancy. To improve the fit, we tried a number of other formulas, obtained by making simple changes to Eq.~\ref{eq:fit1}. In all our attempts, one point at $220\unit{K}$ and $920\unit{kg\,m^{-3}}$, at the corner of the validity region, caused too large deviations, resulting in a reduced $\chi^2=2.21$ for $\eta$ and $11.4$ for $D$ for our best fit with a modified equation. Yet this state point was well equilibrated, as we checked by performing a $880\unit{ns}$-long simulation run. To keep the change to Eq.~\ref{eq:fit1} to a minimum, we decided to discard this problematic point. We kept all other points in the region of validity of the two-state model~\cite{Biddle_twostructure_2017} to form a second set of simulation data, set 2.

We were able to improve the fit to set 2 by adding a volume term $\DvLDS$ in the activation energy for the LDS (similar to $\DvHDS$ for the HDS), namely:
\begin{eqnarray}
&A(T,P) = A_0 \left( \frac{T}{\Tref} \right)^\nu \nonumber \\
&\exp \left\{\epsilon \left[ \left[1 -f(T,P) \right] \frac{\EHDS + \DvHDS P}{\kb (T- \TB)} \right. \right.\nonumber \\
&+ f(T,P) \left. \left. \frac{\ELDS + \DvLDS P}{\kb T} \right]\right\} \; .
\label{eq:fit2}
\end{eqnarray}
An advantage of Eq.~\ref{eq:fit2} over Eq.~\ref{eq:fit1} is that the former is able to yield a second extremum at low density. This can be understood by studying the derivative of $\ln A$ with respect to pressure:
\begin{eqnarray}
& \left( \frac{\partial \ln A}{\partial P} \right)_T =  \epsilon \left\{ \left[1 -f(T,P) \right] \frac{\DvHDS}{\kb (T- \TB)} \right. \nonumber \\
&+ \left. f(T,P) \frac{\DvLDS}{\kb T} \right. \nonumber \\
& + \left. \left( \frac{\partial f}{\partial P} \right)_T
\left[\frac{\ELDS + \DvLDS P}{\kb T} - \frac{\EHDS + \DvHDS P}{\kb (T- \TB)} \right] \right\} \; .
\label{eq:derivP}
\end{eqnarray}
At high pressure, $f\rightarrow 0$ and $(\partial f/\partial P)_T \rightarrow 0$, so that the dynamic behavior is normal, tending towards that of a pure HDS liquid. At intermediate pressures, the $(\partial f/\partial P)_T$ term has a sign opposite to the others, and, if its amplitude is sufficient (i.e. at low enough temperature), it causes the anomalous behavior of dynamic properties. When the pressure is sufficiently reduced, the $1-f$ term can dominate, causing the dynamic properties to recover a normal behavior.

The fit to Eq.~\ref{eq:fit2} and the corresponding residuals are shown in Fig.~\ref{fig:simulfit2}. The fit is good, with significantly better quality than the fit of set 1 to Eq.~\ref{eq:fit1}. The residuals are reasonable, although some bias remains at low temperature and at the two lowest densities. There are several possible reasons for this discrepancy, and for our need to discard the point at $220\unit{K}$ and $920\unit{kg\,m^{-3}}$. The simple linear pressure dependence of the apparent activation energies in Eq.~\ref{eq:fit2} might not be sufficient for the large pressure range investigated; or some parameters of the thermodynamic two-state model (e.g. the location of the Widom line) might have to be modified, to improve the agreement with the dynamic data, without deteriorating the description of thermodynamic data. A simultaneous fit of both types of data is an interesting direction for future work.

For comparison, we also performed the fit of experimental data to Eq.~\ref{eq:fit2}, as shown in Fig.~\ref{fig:expfit2}. Table~\ref{tab:bestfit2} gives the best fit parameters. Adding the $\DvLDS$ term also improves the fit to experiment, albeit only slightly, presumably because of the restricted pressure interval and small values of the LDS fraction in the experimentally covered range. The values of $\ELDS$, $\DvHDS$, $\EHDS$, and $\DvHDS$ are in the same range for the different dynamic quantities. $\ELDS/\kB$, in the range $2000$--$2600\unit{K}$, still has the order of the energy of a hydrogen bond, whereas $\DvHDS$, $\EHDS$, and $\DvHDS$ are more different between simulations and experiment. $\TB$ is nearly the same as for the previous fit of the experimental data, whereas it is increased to $159\unit{K}$ for the fit of MD data. The activation volume $\DvHDS$ is slightly increased but remains small, while the activation volume $\DvLDS$ is rather large, in the range $15$--$42\,10^{-30}\unit{m^3}$. This value is similar to the volume per molecule in the liquid. In the model we propose, transport by a molecule in the LDS state would thus involve a considerable change in volume for the activated state. This is not unlikely, as the LDS state is sometimes viewed as a structure involving a tetrahedral arrangement of hydrogen bonded molecules, with low entropy and large volume.

\begin{figure}[ttt]
\centering
\centerline{\includegraphics[width=.85\columnwidth]{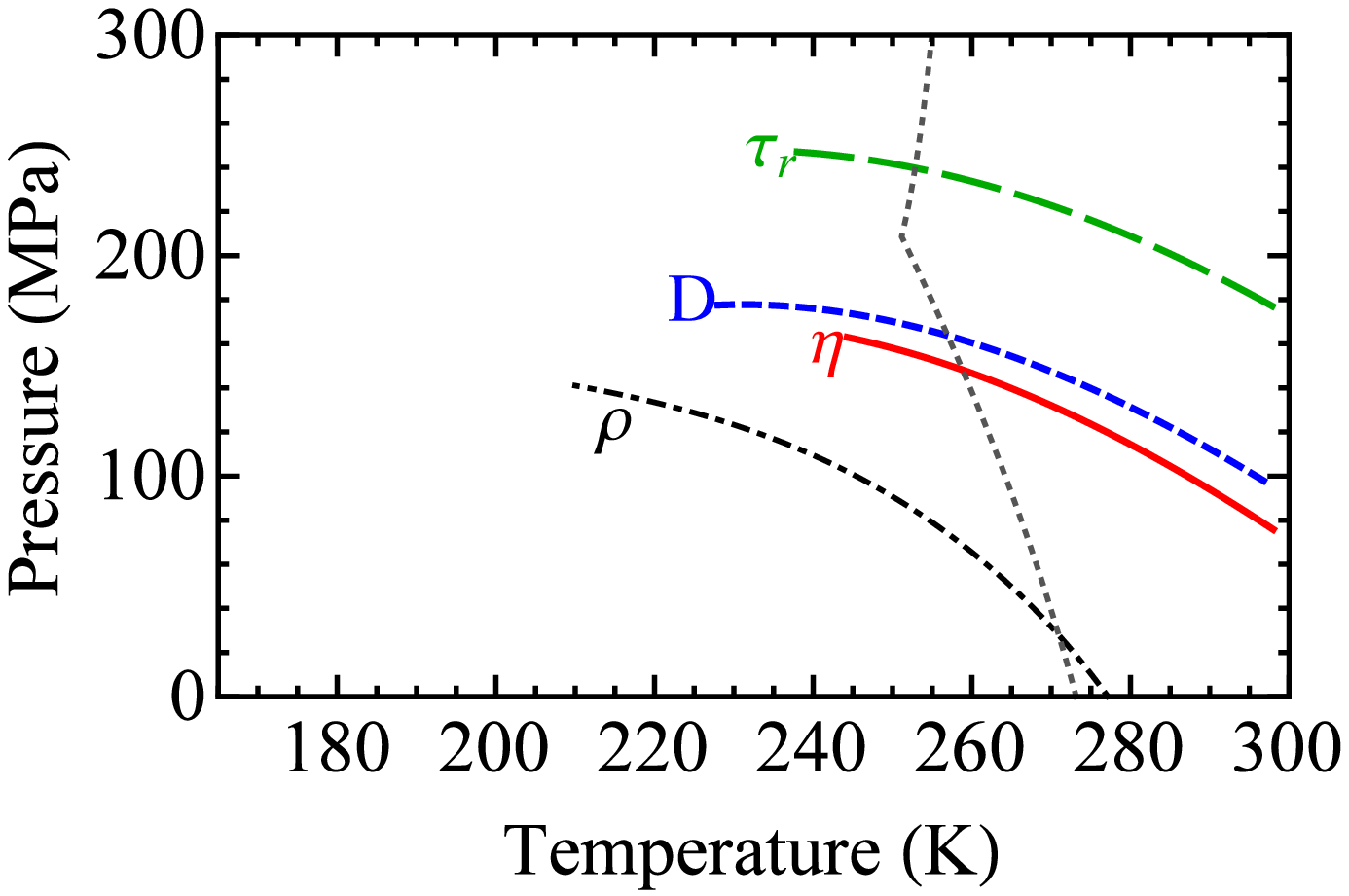}}
\centerline{\includegraphics[width=.85\columnwidth]{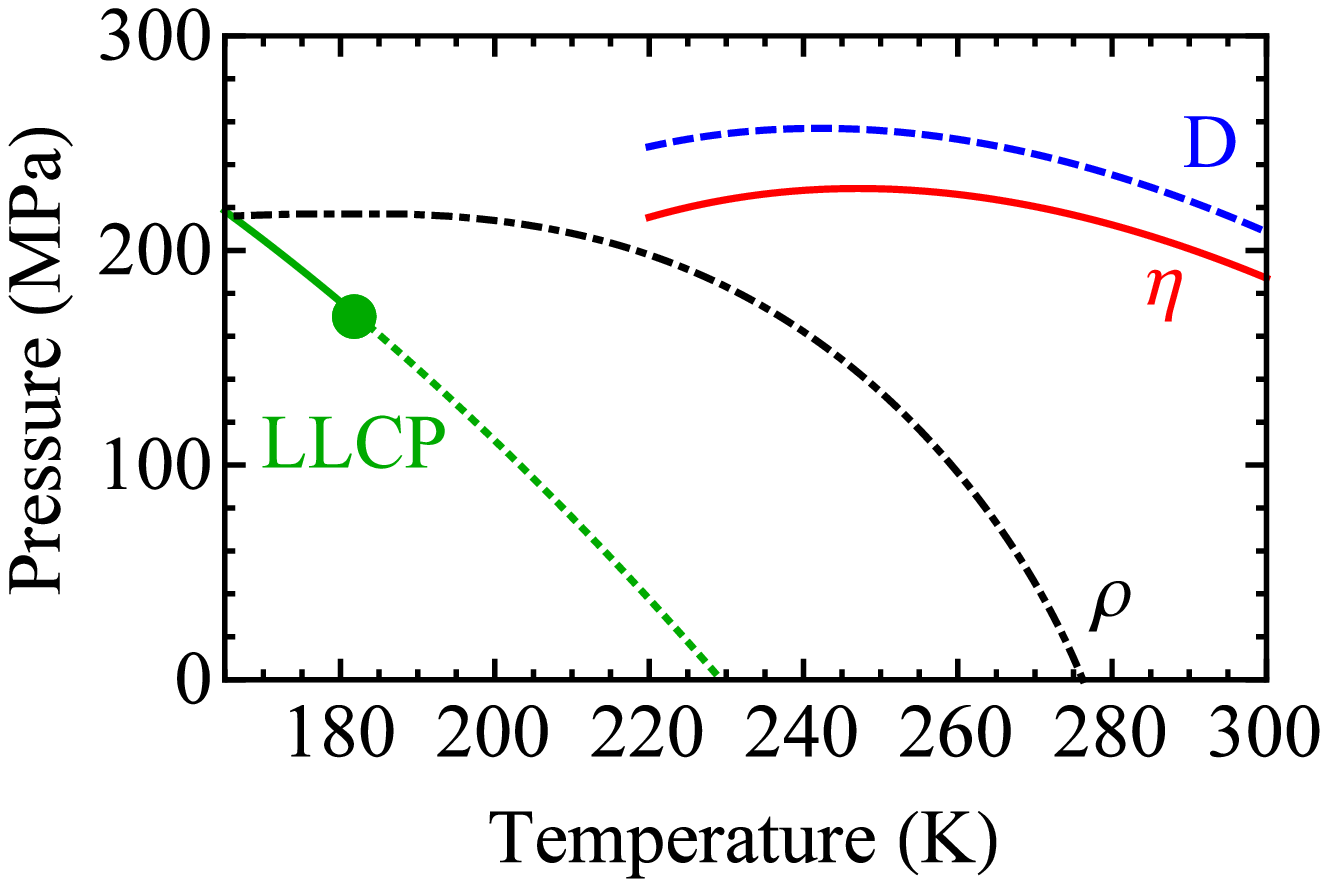}}
\caption{\textbf{Extrema in density and dynamic properties.} Top: Location in the pressure-temperature plane of the experimental extrema along isotherms for viscosity $\eta$ (full red curve), self-diffusion coefficient $D$ (short-dashed blue curve), rotational correlation time $\taur$ (long-dashed green curve), and density $\rho$ (dash-dotted black curve). The gray dotted curve shows the melting lines of ice Ih and ice III~\cite{Wagner_new_2011}. Adapted from Ref.~\onlinecite{Singh_pressure_2017}, where experiments were fit with Eq.~\ref{eq:fit1}. Bottom: same as top, but for the TIP4P/2005 model and Eq.~\ref{eq:fit1}, and including the liquid-liquid transition (solid green line), the LLCP, and the Widom line (dotted green line)~\cite{Biddle_twostructure_2017}.
\label{fig:linesfit1}}
\end{figure}

\begin{figure}[ttt]
\centering
\centerline{\includegraphics[width=.85\columnwidth]{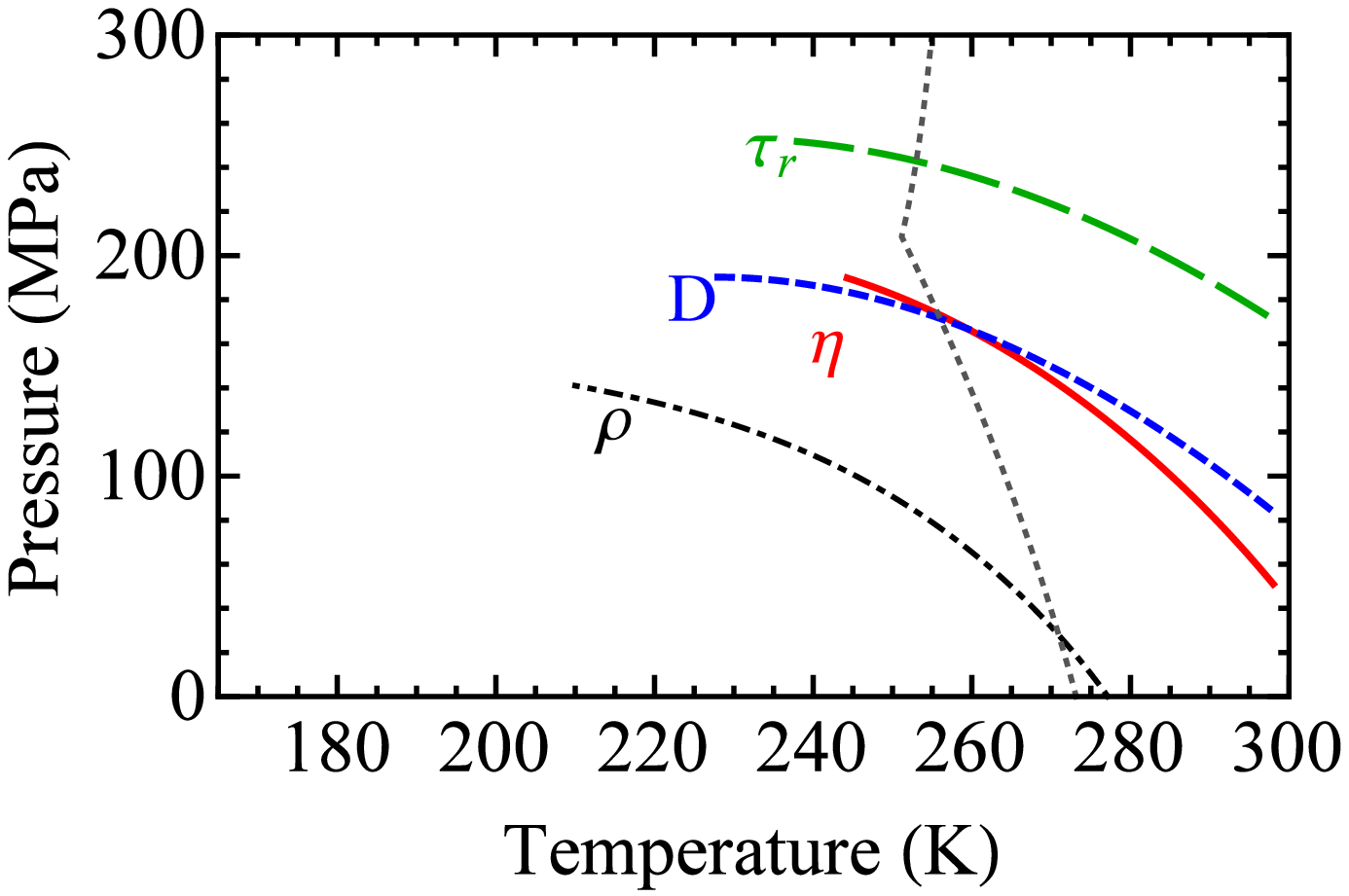}}
\centerline{\includegraphics[width=.85\columnwidth]{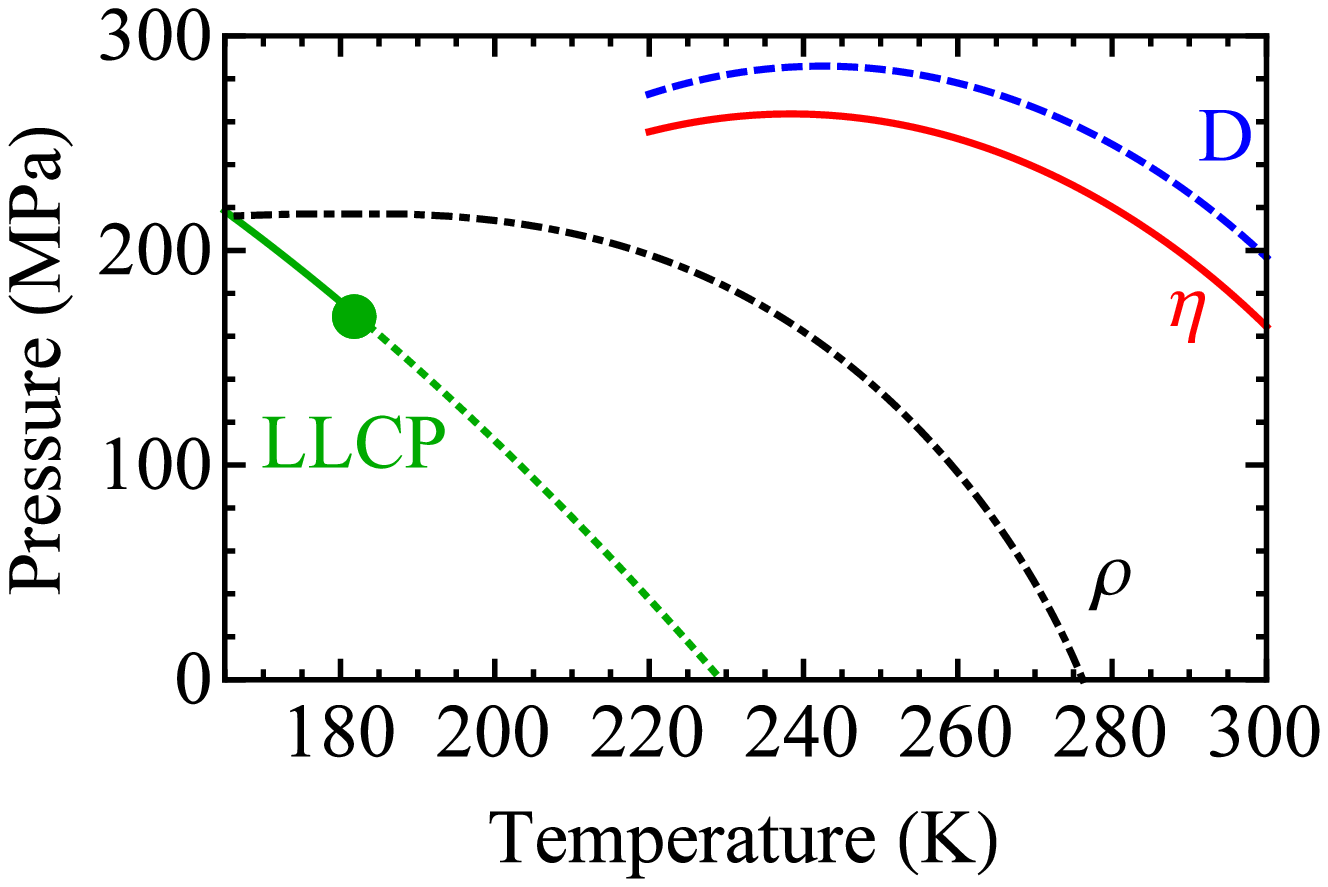}}
\caption{\textbf{Extrema in density and dynamic properties.} Same as previous figure, except that the experimental dynamic data (top) were fit with Eq.~\ref{eq:fit2} (Fig.~\ref{fig:expfit2} and Table~\ref{tab:bestfit2}), and the TIP4P/2005 simulation set 2 (bottom) were fit with Eq.~\ref{eq:fit2} (Fig.~\ref{fig:simulfit2} and Table~\ref{tab:bestfit2}).
\label{fig:linesfit2}}
\end{figure}

We now discuss the value of $\TB$ appearing in the VTF-like behavior of the dynamics of the HDS state, Eqs.~\ref{eq:fit1} and ~\ref{eq:fit2}. $\TB$, at which the system would be arrested, has been related to the Kauzmann temperature~\cite{Adam_temperature_1965} or the mode-coupling temperature~\cite{Gotze_essentials_1998}. In the former case, it is expected to be lower than $\Tg$, whereas in the latter case, $\TB$ should be higher than $\Tg$, because of hopping processes. $\Tg$ for water has been reported below $145\unit{K}$~\cite{Loerting_glass_2015}. However, a recent comparison of the calorimetric features of the glass phases of several water isotopes~\cite{Shephard_molecular_2016} points towards a reinterpretation of the glass transition as an orientational glass transition. The true structural glass transition of water might therefore occur at temperature above $145\unit{K}$, which eludes observation because of crystallization upon further heating. We make the conservative statement that the best fit value for $T_0$ is close to $\Tg$.

Finally, we compare the lines of extrema for experiment and simulations, as derived from the fitting of experiment and simulation set 1 to Eq.~\ref{eq:fit1} (Fig.~\ref{fig:linesfit1}), and of experiment and simulation set 2 to Eq.~\ref{eq:fit2} (Fig.~\ref{fig:linesfit2}). The line of density maxima is also shown, together with the liquid-liquid transition and the Widom line for TIP4P/2005. All figures are qualitatively similar. We note that Fig.~\ref{fig:linesfit1} does not show the intersection between the line of minima in $\eta$ and of maxima in $D$ for the fit to experiment, nor the maxima of these lines for the fit to simulations, which can be seen in Fig.~\ref{fig:linesfit2}. We believe that these features are not significant, and rather due to inaccuracies of the fit in locating the rather shallow extrema (see Figs.~\ref{fig:simulfit1} to \ref{fig:expfit2}). A robust result is the nested pattern formed by the lines. Part of this pattern was observed in previous simulations~\cite{Errington_relationship_2001,Poole_dynamical_2011,Nayar_water_2013}, with the locus of maxima in $D$ encircling the line of density maxima. The same arrangement of these lines was also observed for mW water, with in addition the locus of minima in $\eta$ located in between them. However, mW does not reproduce quantitatively the dynamics of real water (see Section~\ref{sec:intro}). Here, with the more quantitative TIP4P/2005 water model, we find the lines of extrema in the same order as, and at a location close to, the experimental lines of extrema.

\subsection{Stokes-Einstein relation\label{sec:SE}}

\begin{figure}[ttt]
\centering
\centerline{\includegraphics[width=0.8\columnwidth]{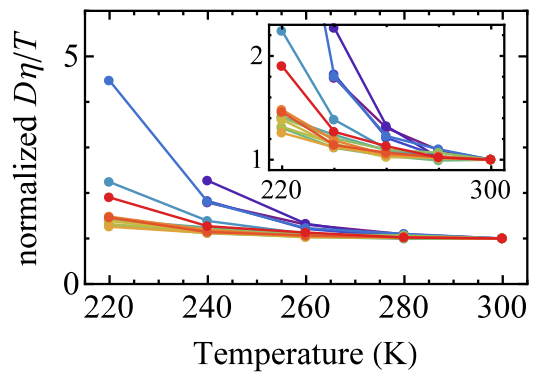}}
\caption{Temperature variation of the quantity $D \eta/T$ normalized by its value at $300\unit{K}$ for a series of isochores. The corresponding densities are listed in Appendix~\ref{sec:data}, and the color code is the same as in Appendix~\ref{sec:Arrhenius}. The lines connecting data points are to guide the eye. The inset shows a zoom to emphasize the non-monotonic density variation.\label{fig:SE_T}}
\end{figure}

We are now in a position to test the SE relation by combining the simulation results. We choose to use directly the raw simulation data rather than the fits presented in Section~\ref{sec:2statesresults}, because the simulations cover a larger range of temperature and pressure. Moreover, in their validity region, the fits exhibit systematic deviations which, although small for the absolute values of $\eta$ and $D$ compared to the simulation uncertainties, result in an excessive underestimate of the product $D \eta$. To emphasize the temperature variation, $D \eta/T$ is usually normalized at a reference temperature, which is taken as $300\unit{K}$ in Fig.~\ref{fig:SE_T}. For $1000\unit{kg\,m^{-3}}$, the violation reaches 24\% at $240\unit{K}$, which is comparable to the violation of around 60\% observed in the experiment at $240\unit{K}$ and atmospheric pressure. At a given temperature, the SE violation tends to become more pronounced at lower densities; however, the density dependence is not monotonic. 

\begin{figure}[ttt]
\centering
\centerline{\includegraphics[width=0.85\columnwidth]{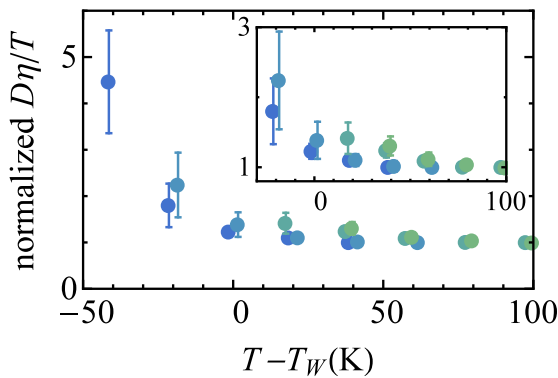}}
\caption{Temperature variation of the quantity $D \eta/T$ normalized by its value at $300\unit{K}$, as a function of the distance to the Widom line $T-T_\mrm{W}(\rho)$ (see text for details) for four isochores (from bottom to top: 920.050, 960.090, 999.260, and $1040.59\unit{kg\,m^3}$). The inset points out the non-perfect collapse of the three isochores.\label{fig:SE_T-TW}}
\end{figure}

Kumar \etal~\cite{Kumar_relation_2007} studied the SE relation for two other models of water: TIP5P and ST2. Note that they used the structural relaxation time $\tau_\alpha$ as a proxy for the shear viscosity $\eta$ (see Section~\ref{sec:intro}). They related the violation of SE to the existence of a LLCP in the supercooled liquid, and more particularly to the Widom line emanating from this LLCP, located at a temperature $T_\mrm{W}$ function of the pressure $P$. They found that, at pressures lower than the LLCP pressure, the $D \tau_\alpha/T$ curves for each pressure collapsed onto a master curve when plotted as a function of the distance to the Widom line, $T-T_\mrm{W}(P)$, instead of the temperature. We have tested this collapse. Strictly speaking, the Widom line is the locus of correlation length maxima associated with the LLCP. As a proxy for $T_\mrm{W}(P)$, Kumar~{\etal} used the maxima of isobaric heat capacity along isobars, which asymptotically approaches the Widom line near the LLCP. Here instead, we use the two-state model presented in Section~\ref{sec:2statesmodel}. For the 4 isochores having a density below the LLCP density, but still in the validy region of the two-state model, we use the two-state model to locate the Widom line as the locus of points where the LDS and HDS have equal fraction, $1/2$. This is given by the roots of Eq.~(4) of Ref.~\onlinecite{Biddle_twostructure_2017}, which correspond to the two states having the same Gibbs free energy. Figure~\ref{fig:SE_T-TW} shows the normalized $D \eta/T$ as a function of $T-T_\mrm{W}(P)$. We observe an approximate collapse, but a density dependence can still be seen.

\begin{figure}[ttt]
\centering
\centerline{\includegraphics[width=0.82\columnwidth]{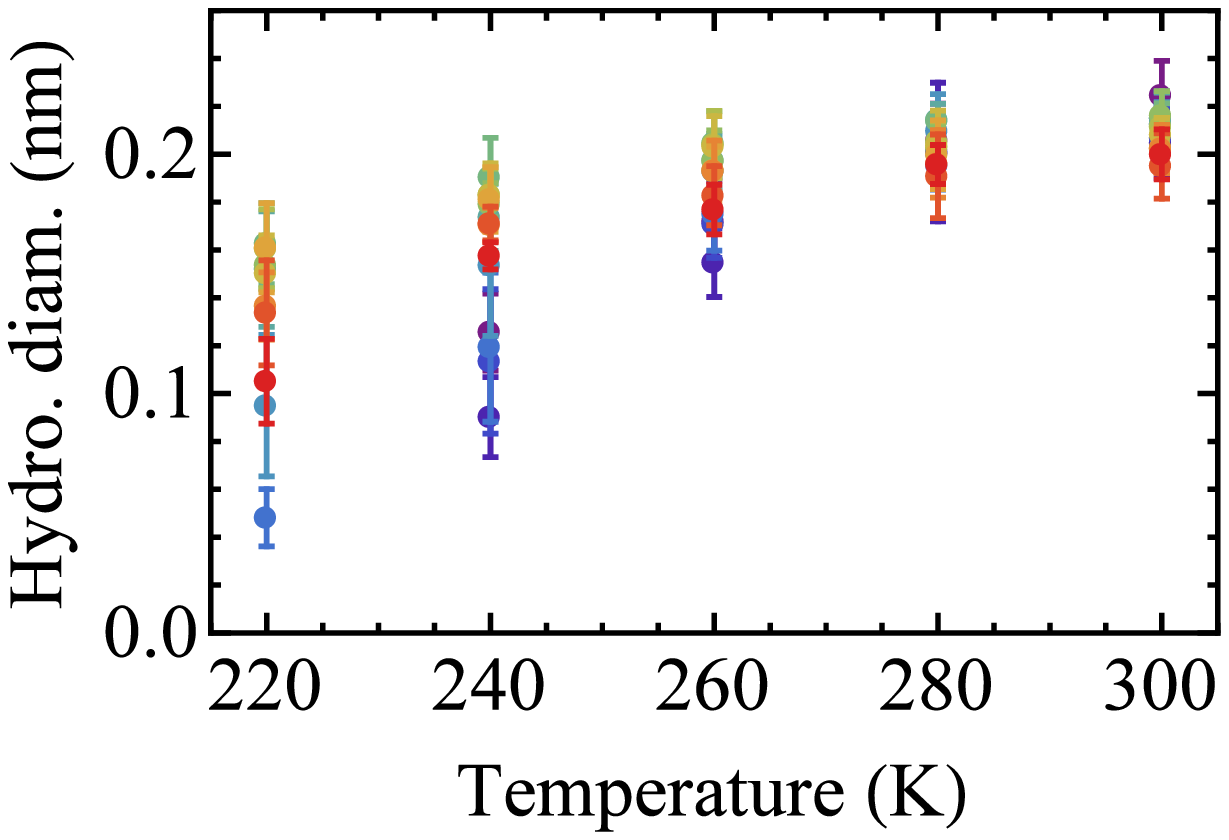}}
\vspace*{2mm}
\centerline{\includegraphics[width=0.82\columnwidth]{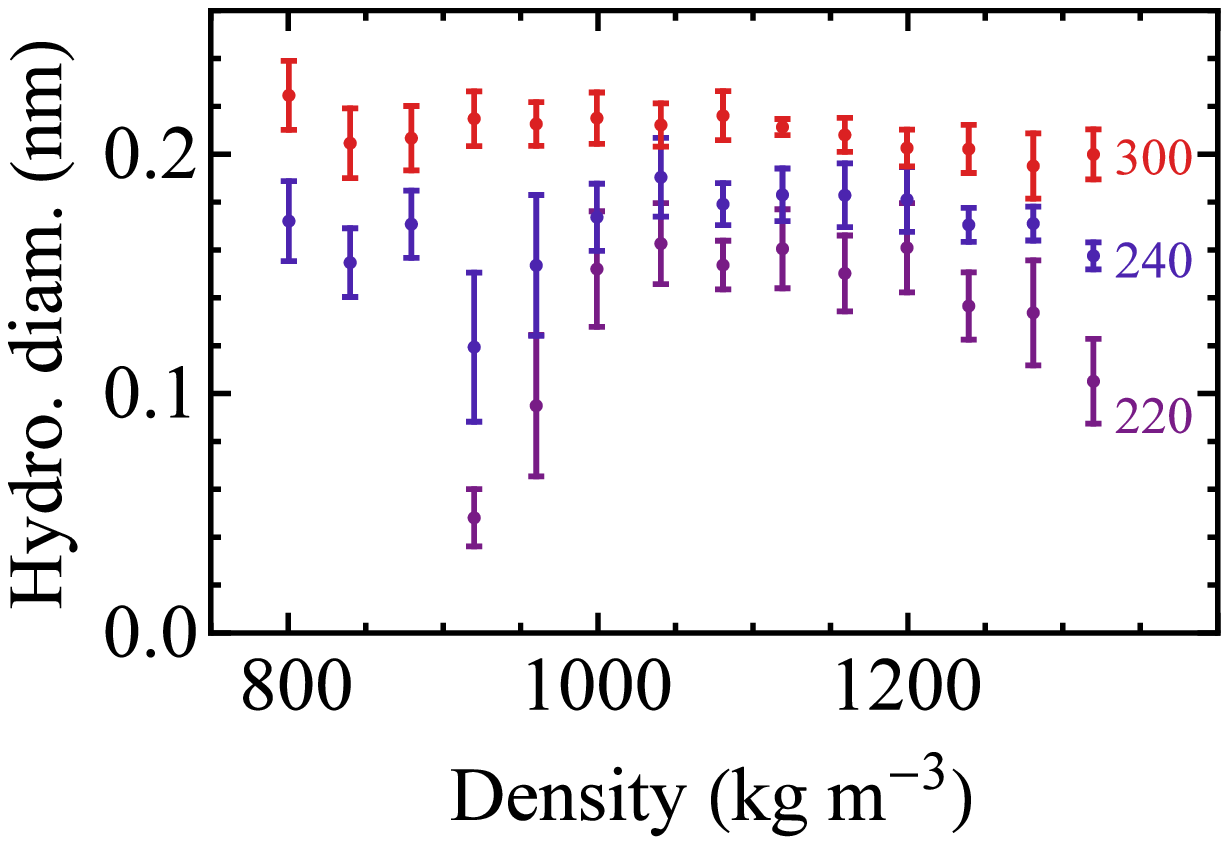}}
\caption{Top: Temperature dependence of the hydrodynamic diameter $\phi_\mrm{h}$ for a series of isochores with the same color code as in Appendix~\ref{sec:Arrhenius}. Bottom: Density dependence of the hydrodynamic diameter $\phi_\mrm{h}$ for three isotherms (labels give the temperature in K).\label{fig:SEdiam}}
\end{figure}

The normalization process used above removes the information about the absolute value of $D \eta/T$. If $D$ was the diffusion coefficient of a macroscopic object obeying hydrodynamics in the Stokes regime, $D \eta/T$ would be related to the hydrodynamic diameter $\phi_\mrm{h}$ by:
\beq
\phi_\mrm{h} = \frac{\kb T}{3 \pi \eta D} \; .
\label{eq:hydro}
\eeq
Figure~\ref{fig:SEdiam} shows $\phi_\mrm{h}$ computed from the simulation data. At high temperature, $\phi_\mrm{h}$ is $0.2$--$0.22\unit{nm}$, nearly independent of (or only slightly decreasing with) density. To assess the validity of Eq.~\ref{eq:hydro}, $\phi_\mrm{h}$ should be compared to a molecular diameter determined independently. Several choices of this molecular diameter are possible (see for instance Ref.~\onlinecite{Cappelezzo_stokeseinstein_2007} for a discussion in the case of the Lennard-Jones fluid). The volume per molecule in the liquid is around $30\,10^{-30}\unit{m^3}$ at $\rho=1000\unit{kg\,m^{-3}}$, equivalent to a sphere of diameter $0.38\unit{nm}$, or $0.33\unit{nm}$ if one considers random close-packed spheres occupying 64\% of space. The Lennard-Jones parameter for interaction between the oxygen sites of two
molecules in TIP4P/2005 is $0.31589\unit{nm}$.~\cite{Abascal_general_2005} All these values are close to $\phi_\mrm{h}$. For a spherical object, a hydrodynamic diameter smaller than the physical diameter can be due to the slip boundary condition between the object and the ambient fluid~\cite{Zwanzig_hydrodynamic_1970}. This can vary the factor in the denominator of Eq.~\ref{eq:hydro} from $3 \pi$ (no slip) to $2 \pi$ (perfect slip). Slip could thus explain the values of $\phi_\mrm{h}$ for water at high temperature~\cite{Sposito_single_1981}. A change in slip boundary conditions may also explain changes in $\phi_\mrm{h}$ up to 50\%, but cannot account for the large decrease at low temperature, which can exceed a factor of 10. An explanation based on slippage only should thus be discarded.

The behavior of water is reminiscent of many glassformers near their glass transition temperature $\Tg$. In this case, the decoupling between $D$ and $\eta$ is due to the emergence of dynamic heterogeneities, that is, transient, spatially correlated regions of particles with high and low mobility~\cite{Tarjus_breakdown_1995,Ediger_perspective_2012}. Emergence of these regions at low temperature gives rise to a distribution of relaxation times broader than at high temperature. Because the different dynamic quantities result from different moments of the distribution, they start decoupling upon cooling. The SE violation in water has also been related to dynamic heterogeneities~\cite{Becker_fractional_2006,Mazza_connection_2007,Kumar_relation_2007,Guillaud_decoupling_2017,Galamba_hydrogenbond_2017,Kawasaki_identifying_2017}
; however, the discussion was based on simulations of $\tau_\alpha$ rather than of $\eta$, and in contrast to usual glassformers for which the most mobile molecules cause the breakdown of the SE relation, all scales of mobility were involved in water. Further studies are needed to better understand the origin of the SE violation in water and its relation with the Widom line.

\section{Conclusion\label{sec:conclu}}

By performing extensive simulations of dynamic properties for the TIP4P/2005 water model, we have been able to reproduce nearly quantitatively all features observed for viscosity and self-diffusion coefficent of real water at temperatures below ambient, including the supercooled region, and in a broad positive pressure range. Our simulations also go beyond the conditions which have been hitherto explored in experiments. At lower temperatures, the minimum in $\eta$ and the maximum in $D$ as a function of density or pressure are found to become even more pronounced. At negative pressure, a maximum in $\eta$ and a minimum in $D$ are observed. The dynamic extension of the thermodynamic two-state model available for TIP4P/2005 is able to accurately reproduce the simulation data. Inclusion of a pressure dependence in the activation energy of the low density state is necessary to fit the negative pressure data, pointing to a large activation volume for the dynamics of this state. The Stokes-Einstein relation is strongly violated as the system is cooled through the Widom line. Our study provides a unifying framework to interpret the thermodynamic and dynamic anomalies of water, and calls for experiments on the dynamics of water at negative pressure.

%

\section*{Acknowledgments}
PM, ES and CV have been funded by grants FIS2013/43209-P, FIS2016-78117-P and FIS2016-78847-P of the MEC and the UCM/Santander 910570 and PR26/16-10B-2. 
PM acknowledges financial support from a FPI PhD fellowship. LJ acknowledges support from Institut Universitaire de France. This work was partially supported by CNRS (France) through a PICS program.

\appendix

\section{Simulation data\label{sec:data}}

Tables~\ref{tab:dataeta} and \ref{tab:dataD} give all the simulation results of this study with their uncertainty (one standard deviation). For viscosity (Table~\ref{tab:dataeta}), the uncertainty is the standard deviation of the five independent Green-Kubo integrals of the auto-correlation function of traceless stress tensor elements~\cite{Chen_are_2009}. For self-diffusion (Table~\ref{tab:dataD}), the uncertainty was less straighforward to obtain, and we proceeded as follows. At each temperature, for one every three densities, we used the block averaging method on one of the trajectories. The selected trajectory was cut into four pieces with equal duration. For each piece, the self-diffusion coefficient for the finite system, $D_\mrm{PBC}$, was calculated from the slope of the mean squared displacement $\langle r^2 \rangle$ in the diffusive regime as explained in Section~\ref{sec:simdetails}. The uncertainty on $D_\mrm{PBC}$ was taken as the standard deviation of the four values thus obtained. Table~\ref{tab:dataD} gives the self-diffusion coefficient $D$ for the infinite liquid, after correction for finite size effects using Eq.~\ref{eq:Dcorr}. The total uncertainty on the corrected $D$ was calculated by propagating the uncertainty on $D_\mrm{PBC}$ and $\eta$. Because the procedure was computationally costly, we applied it at every temperature, but only for one every three densities. At each temperature, for each remaining density, we assumed that the relative uncertainty on $D$ was equal to the relative uncertainty on $D$ at the nearest density for which it was directly calculated with the above method. Hence, absolute uncertainties on $D$ at the remaining densities were only calculated indirectly.

We note that, in order to get a more accurate estimate of the uncertainties, more simulations would be needed. The quantity $\chi^2$ we use to assess the quality of the fits is quite sensitive to the uncertainty, because it involves dividing by the squared uncertainties. Therefore the absolute values for $\chi^2$ could be modified if the uncertainty calculations were refined. Nevertheless, because fitting with the original or the modified two-state model uses the same definitions for the uncertainties, the comparison between the two fits is justified. Our results show that the modified model gives a better fit than the original one, and over a broader pressure range.

\begin{table*}[t]
\centering
\caption{\label{tab:dataeta} Simulation results for the shear viscosity $\eta$ in $\mrm{mPa\,s}$. The uncertainty (one standard deviation) is given between parentheses.}
\begin{tabular}{ccccccccccc}
\hline
\hline
	& \multicolumn{5}{c}{Temperature ($\mrm{K}$)}\\
Density ($\mrm{kg\,m^{-3}}$)	& 220	& 240	& 260	& 280	& 300	\\
\hline
800.43&	 &	37.5 (3.8)&	5.72 (0.45)&	1.73 (0.12)&	0.811 (0.046)\\
839.99&	&	101 (17)&	8.46 (0.61)&	2.12 (0.30)&	0.961 (0.062)\\
879.49&	&	82 (17)&	8.26 (0.59)&	2.157 (0.077)&	1.011 (0.062)\\
920.05&	1348 (281)&	45.3 (9.0)&	5.58 (0.43)&	1.929 (0.089)&	0.926 (0.045)\\
960.09&	164 (46)&	15.9 (1.5)&	3.82 (0.25)&	1.57 (0.11)&	0.864 (0.032)\\
999.26&	36.8 (5.7)&	7.94 (0.56)&	2.753 (0.091)&	1.320 (0.037)&	0.834 (0.036)\\
1040.59&	20.3 (2.0)&	5.40 (0.41)&	2.20 (0.10)&	1.214 (0.014)&	0.808 (0.028)\\
1080.66&	15.5 (0.9)&	4.64 (0.13)&	2.114 (0.097)&	1.215 (0.056)&	0.816 (0.033)\\
1119.05&	13.8 (1.3)&	4.24 (0.15)&	2.04 (0.12)&	1.228 (0.028)&	0.8368 (0.0092)\\
1159.29&	14.4 (1.4)&	4.40 (0.24)&	2.15 (0.11)&	1.327 (0.084)&	0.933 (0.030)\\
1199.42&	14.8 (1.6)&	4.83 (0.28)&	2.44 (0.09)&	1.480 (0.093)&	1.029 (0.037)\\
1239.28&	22.6 (1.7)&	6.23 (0.18)&	2.84 (0.11)&	1.71 (0.12)&	1.189 (0.055)\\
1280.93&	33.7 (5.0)&	8.21 (0.23)&	3.54 (0.15)&	2.11 (0.19)&	1.48 (0.10)\\
1319.79&	78.2 (12)&	12.81 (0.24)&	5.02 (0.12)&	2.61 (0.10)&	1.770 (0.087)\\
\hline
\hline
\end{tabular}
\end{table*}

\begin{table*}[t]
\centering
\caption{\label{tab:dataD} Simulation results for the self-diffusion coefficient $D$ in $10^{-9}\unit{m^2\,s^{-1}}$ after correction with Eq.~\ref{eq:Dcorr}. The uncertainty (one standard deviation) is given between parentheses. Uncertainty values in italics were calculated from values for neighboring densities.}
\begin{tabular}{ccccccccccc}
\hline
\hline
	& \multicolumn{5}{c}{Temperature ($\mrm{K}$)}\\
Density ($\mrm{kg\,m^{-3}}$)	& 220	& 240	& 260	& 280	& 300	\\
\hline
800.43&	&	0.0746 (0.0058)&	0.387 (0.022)&	1.152 (0.031)&	2.413 (0.072)\\
839.99&	&	0.0386 (\textit{0.0030})&	0.291 (\textit{0.017})&	0.963 (\textit{0.026})&	2.235 (\textit{0.067})\\
879.49&	&	0.0378 (\textit{0.0063})&	0.270 (\textit{0.011})&	0.949 (\textit{0.021})&	2.103 (\textit{0.046})\\
920.05&	0.00497 (0.00067)&	0.065 (0.011)&	0.390 (0.016)&	1.089 (0.024)&	2.209 (0.048)\\
960.09&	0.0207 (\textit{0.0028})&	0.144 (\textit{0.024})&	0.517 (\textit{0.021})&	1.246 (\textit{0.028})&	2.392 (\textit{0.052})\\
999.26&	0.0576 (\textit{0.0020})&	0.255 (\textit{0.010})&	0.701 (\textit{0.032})&	1.450 (\textit{0.021})&	2.450 (\textit{0.061})\\
1040.59&	0.0976 (0.0033)&	0.342 (0.014)&	0.846 (0.039)&	1.579 (0.023)&	2.563 (0.064)\\
1080.66&	0.1353 (\textit{0.0046})&	0.423 (\textit{0.017})&	0.914 (\textit{0.042})&	1.666 (\textit{0.025})&	2.492 (\textit{0.062})\\
1119.05&	0.1455 (\textit{0.0060})&	0.453 (\textit{0.022})&	0.915 (\textit{0.033})&	1.627 (\textit{0.056})&	2.485 (\textit{0.029})\\
1159.29&	0.1490 (0.0061)&	0.437 (0.021)&	0.872 (0.032)&	1.519 (0.053)&	2.264 (0.026)\\
1199.43&	0.1353 (\textit{0.0056})&	0.402 (\textit{0.019})&	0.808 (\textit{0.029})&	1.387 (\textit{0.048})&	2.108 (\textit{0.024})\\
1239.28&	0.1044 (\textit{0.0073})&	0.331 (\textit{0.010})&	0.695 (\textit{0.037})&	1.224 (\textit{0.020})&	1.828 (\textit{0.033})\\
1280.93&	0.0715 (0.0050)&	0.2504 (0.0076)&	0.589 (0.031)&	1.019 (0.017)&	1.522 (0.027)\\
1319.79&	0.0392 (\textit{0.0027})&	0.1742 (\textit{0.0053})&	0.429 (\textit{0.023})&	0.803 (\textit{0.013})&	1.242 (\textit{0.022})\\
\hline
\hline
\end{tabular}
\end{table*}

\section{Arrhenius plots\label{sec:Arrhenius}}

\begin{figure*}[ttt]
\centering
\centerline{\includegraphics[width=0.72\columnwidth]{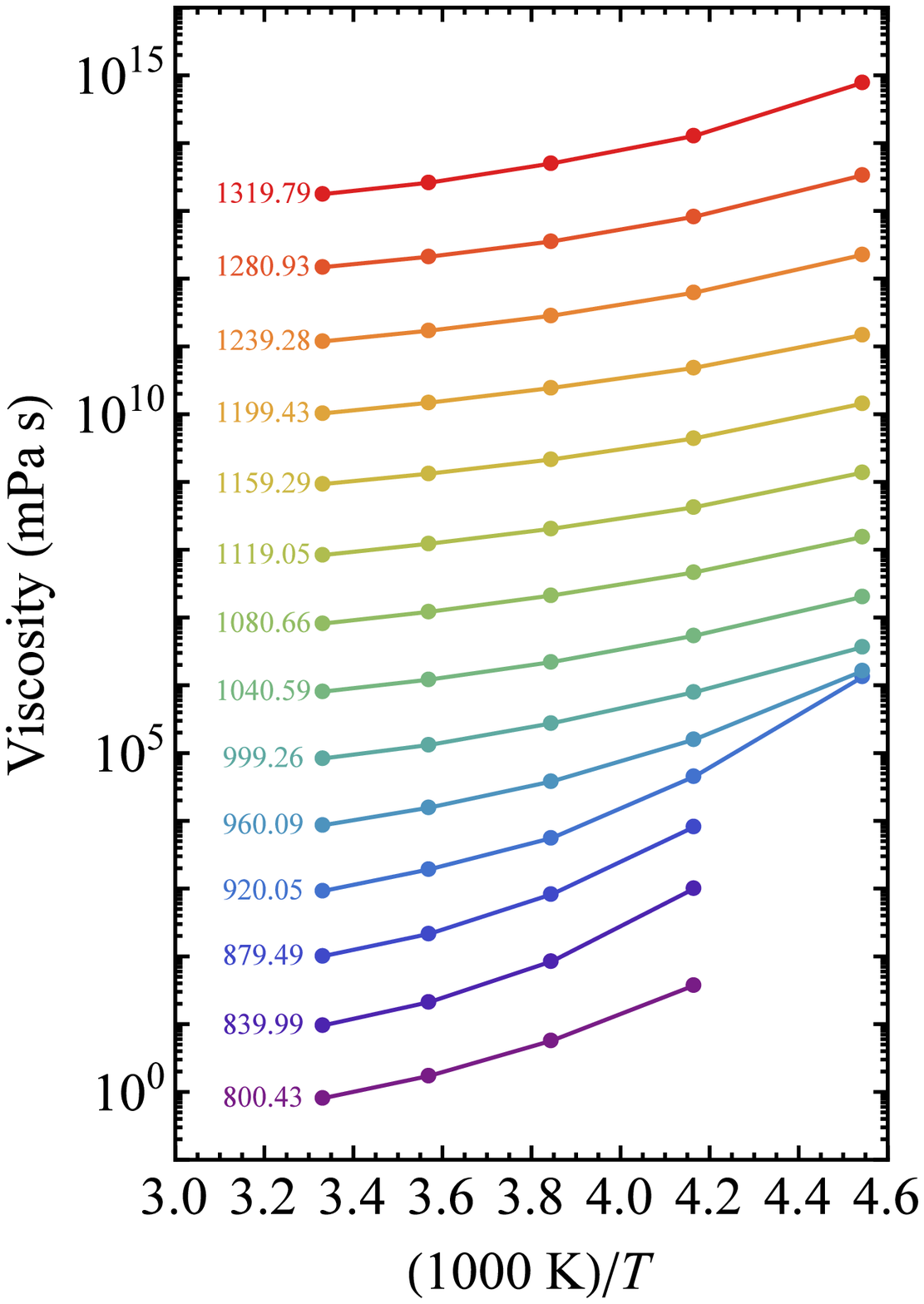}\hspace{5mm}\includegraphics[width=0.72\columnwidth]{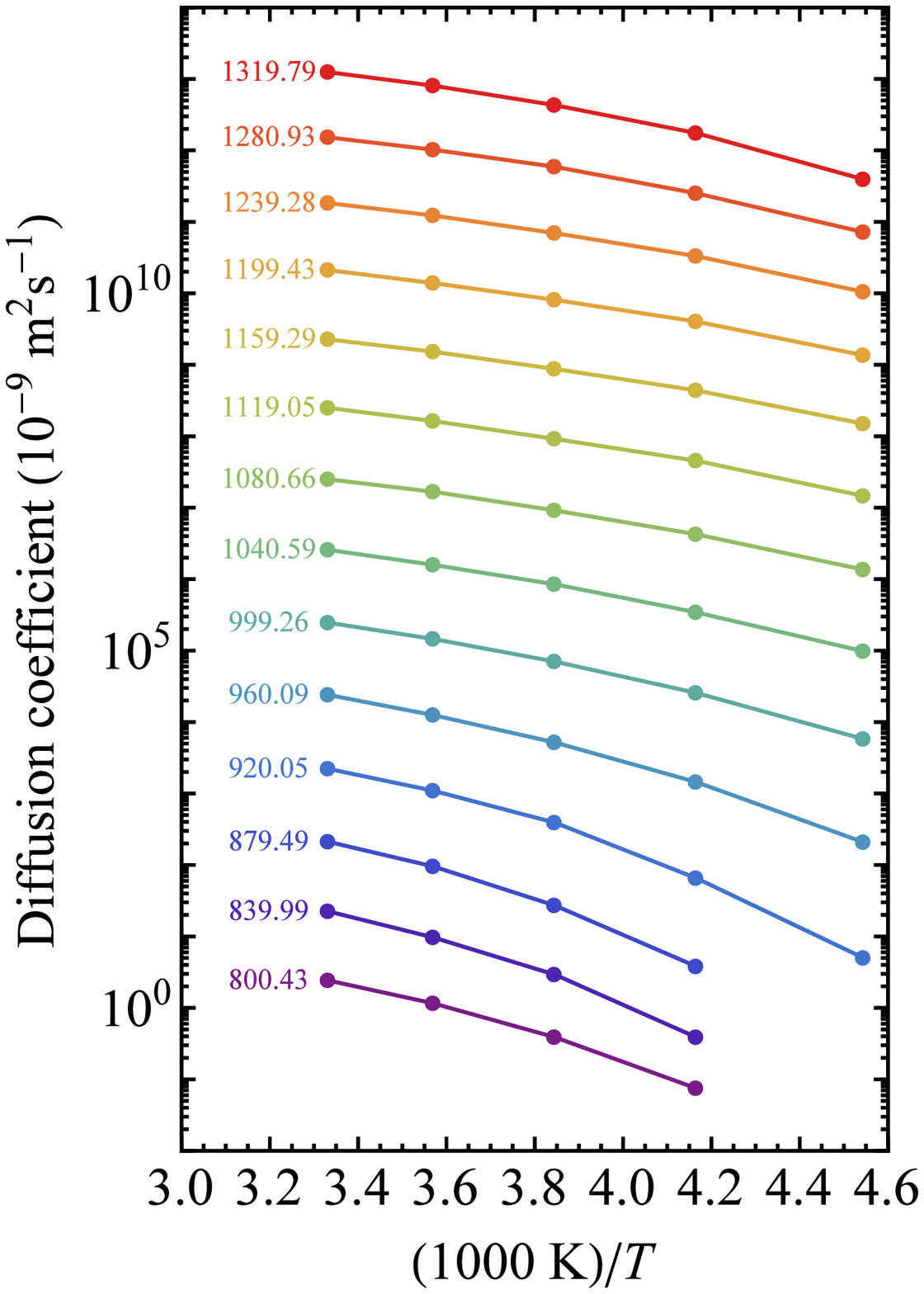}}
\caption{Arrhenius plots for shear viscosity (left) and self-diffusion coefficient (right) for a series of isochores, labeled by the density in $\unit{kg\,m^{-3}}$. The data sets have been successively multiplied by 10 for clarity, with lines connecting points to guide the eye.\label{fig:Arrhenius}}
\end{figure*}

Figure~\ref{fig:Arrhenius} gives a log-lin plot of $\eta$ and $D$ vs. inverse temperature.

\bibliography{articles}

\end{document}